\documentclass{emulateapj}

\usepackage{natbib}
\usepackage{hyperref}
\usepackage{color}
\usepackage{xspace}

\newcommand{\boldsymbol}[1]{\mbox{\boldmath{${#1}$}}}

\shorttitle{Dark matter content and stellar IMF of massive early-type galaxies}
\shortauthors{Sonnenfeld et~al.}

\def\ucsb{1}
\def\ucla{2}
\def\kipac{3}
\def\asiaa{4}
\def\iap{5}
\def\cambridge{6}
\def\bologna{7}

\def\pr{{\rm Pr}}

\def\Sref#1{Section~\ref{#1}\xspace}
\def\Fref#1{Figure~\ref{#1}\xspace}
\def\Tref#1{Table~\ref{#1}\xspace}
\def\Eref#1{Equation~\ref{#1}\xspace}

\def\gammadm{\gamma_{\mathrm{DM}}}
\def\gammadmi{\gamma_{\mathrm{DM},i}}
\def\mdm{M_{\mathrm{DM}5}}
\def\mdme{M_{\mathrm{DM}e}}
\def\mdmi{M_{\mathrm{DM}5,i}}
\def\mstari{M_{*,i}}
\def\msps{M_*^{\mathrm{(SPS)}}}
\def\mspsi{M_{*,i}^{\mathrm{(SPS)}}}
\def\reff{R_{\mathrm{e}}}
\def\reffi{R_{\mathrm{e},i}}
\def\rein{R_{\mathrm{Ein}}}
\def\reini{R_{\mathrm{Ein},i}}
\def\aimf{\alpha_{\mathrm{IMF}}}
\def\aimfi{\alpha_{\mathrm{IMF},i}}

\def\fdm{f_{\mathrm{DM}e}}

\def\hyperp{\boldsymbol{\tau}}
\def\indhyperp{\boldsymbol{\psi}}
\def\dephyperp{\boldsymbol{\theta}}
\def\selhyperp{\boldsymbol{\lambda}}
\def\individ{\boldsymbol{\omega}_i}
\def\indind{\boldsymbol{\xi}_i}
\def\depind{\boldsymbol{\eta}_i}
\def\datad{\mathbf{d}}
\def\datai{\mathbf{d}_i}

\def\mmu{\mu_*}
\def\mz{\zeta_*}
\def\mpiv{\mu_{*,0}}
\def\msig{\sigma_*}
\def\rmu{\mu_R}
\def\rz{\zeta_R}
\def\rmstar{\beta_R}
\def\rpiv{\mu_{R,0}}
\def\rsig{\sigma_R}

\def\mmusl2s{\mu_*^{(\mathrm{SL2S})}}
\def\mzsl2s{\zeta_*^{(\mathrm{SL2S})}}
\def\mpivsl2s{\mu_{*,0}^{(\mathrm{SL2S})}}
\def\msigsl2s{\sigma_*^{(\mathrm{SL2S})}}
\def\rmusl2s{\mu_R^{(\mathrm{SL2S})}}
\def\rzsl2s{\zeta_R^{(\mathrm{SL2S})}}
\def\rmsl2s{\beta_R^{(\mathrm{SL2S})}}
\def\rpivsl2s{\mu_{R,0}^{(\mathrm{SL2S})}}
\def\rsigsl2s{\sigma_R^{(\mathrm{SL2S})}}

\def\mmuslacs{\mu_*^{(\mathrm{SLACS})}}
\def\mzslacs{\zeta_*^{(\mathrm{SLACS})}}
\def\mpivslacs{\mu_{*,0}^{(\mathrm{SLACS})}}
\def\msigslacs{\sigma_*^{(\mathrm{SLACS})}}
\def\rmuslacs{\mu_R^{(\mathrm{SLACS})}}
\def\rzslacs{\zeta_R^{(\mathrm{SLACS})}}
\def\rmslacs{\beta_R^{(\mathrm{SLACS})}}
\def\rpivslacs{\mu_{R,0}^{(\mathrm{SLACS})}}
\def\rsigslacs{\sigma_R^{(\mathrm{SLACS})}}

\def\mdmmu{\mu_{\mathrm{DM}}}
\def\mdmz{\zeta_{\mathrm{DM}}}
\def\mdmm{\beta_{\mathrm{DM}}}
\def\mdms{\xi_{\mathrm{DM}}}
\def\mdmsig{\sigma_{\mathrm{DM}}}
\def\mdmpiv{M_{\mathrm{DM},0}}

\def\gammamu{\mu_\gamma}

\def\gammasig{\sigma_\gamma}
\def\gammapiv{\gamma_0}

\def\aimfmu{\mu_{\mathrm{IMF}}}
\def\aimfz{\zeta_{\mathrm{IMF}}}
\def\aimfm{\beta_{\mathrm{IMF}}}
\def\aimfs{\xi_{\mathrm{IMF}}}
\def\aimfsig{\sigma_{\mathrm{IMF}}}
\def\aimfpiv{\alpha_{\mathrm{IMF},0}}

\def\scross{X_{\mathrm{lens}}}

\def\example{SL2SJ142059+563007}

\begin{document}

\title{The SL2S Galaxy-scale Lens Sample.  V. 
Dark matter halos and stellar IMF of massive Early-type Galaxies out to redshift 0.8}

\author{Alessandro~Sonnenfeld\altaffilmark{\ucsb}$^{*}$}
\author{Tommaso~Treu\altaffilmark{\ucsb,\ucla}$^{\dag}$}
\author{Philip~J.~Marshall\altaffilmark{\kipac}}
\author{Sherry~H.~Suyu\altaffilmark{\asiaa}}
\author{Rapha\"el~Gavazzi\altaffilmark{\iap}}
\author{Matthew~W.~Auger\altaffilmark{\cambridge}}
\author{Carlo~Nipoti\altaffilmark{\bologna}}

\altaffiltext{\ucsb}{Physics Department, University of California, Santa Barbara, CA 93106, USA} 
\altaffiltext{\ucla}{Department of Physics and Astronomy, University of California, Los Angeles, CA 90025-1547, USA}
\altaffiltext{\kipac}{Kavli Institute for Particle Astrophysics and Cosmology, P.O.~Box 20450, MS29, Stanford, CA 94309, USA}
\altaffiltext{\asiaa}{Institute of Astronomy and Astrophysics, Academia Sinica, P.O.~Box 23-141, Taipei 10617, Taiwan}
\altaffiltext{\iap}{Institut d'Astrophysique de Paris, UMR7095 CNRS - Universit\'e Pierre et Marie Curie, 98bis bd Arago, 75014 Paris, France}
\altaffiltext{\cambridge}{Institute of Astronomy, University of Cambridge, Madingley Rd, Cambridge CB3 0HA, UK}
\altaffiltext{\bologna}{Department of Physics and Astronomy, Bologna University, viale Berti-Pichat 6/2, I-40127 Bologna, Italy}

\altaffiltext{*}{{\tt sonnen@physics.ucsb.edu}}
\altaffiltext{$\dag$}{{Packard Research Fellow}}


\begin{abstract}

We investigate the cosmic evolution of the internal structure of
massive early-type galaxies over half of the age of the Universe. 
We perform a joint lensing and stellar dynamics analysis of a sample of 81 strong lenses from the SL2S and SLACS surveys and combine the results with a hierarchical Bayesian inference method to measure the distribution of dark matter mass and stellar IMF across the population of massive early-type galaxies.
Lensing selection effects are taken into account.
We find that the dark matter mass projected within the inner 5 kpc increases for increasing redshift, decreases for increasing stellar mass density, but is roughly constant along the evolutionary tracks of early-type galaxies.
The average dark matter slope is consistent with that of an NFW profile, but is not well constrained.
The stellar IMF normalization is close to a Salpeter IMF at $\log{M_*} = 11.5$ and scales strongly with increasing stellar mass. No dependence of the IMF on redshift or stellar mass density is detected.
The anti-correlation between dark matter mass and stellar mass density supports the idea of mergers being more frequent in more massive dark matter halos.

\end{abstract}



\section{Introduction}\label{sect:intro}

Early-type galaxies (ETGs) constitute a family of objects of remarkable regularity, captured by tight scaling relations such as the fundamental plane \citep{Dre++87,D+D87} and the relations between central black hole mass and galaxy properties \citep{F+M00,Geb++00,M+H03,H+R04}.
Despite tremendous efforts, it is still unknown what the fundamental source of this regularity is.
Numerical simulations are now able to reproduce some of the key observables of ETGs \citep{Hop++09e,Dub++13,Rem++13,F+M14}, but the resolution and statistics that can be reached today are still too low to allow for meaningful quantitative tests. 
It is still very challenging to obtain realistic simulations of the baryonic component of ETGs, since this is affected by a number of complex physical processes including star formation, feedback from supernovae and the effect of an active galactic nucleus (AGN).
Semi-analytical studies of the evolution of ETGs based on dissipationless mergers seem to be able to match the observed mean size-mass relation, but there are difficulties in matching the observed scatter \citep{Nip++12,Sha++13}.
In addition, dissipationless mergers appear not to be able to reproduce the evolution (or lack thereof) in the total density profile of massive ETGs \citep{SNT14}.
\citet{Por++14} suggest that dissipational effects are critical for correctly predicting the normalization and scatter of the fundamental plane relation, but further tests are needed to check whether their model matches the entire set of observations of ETGs.
On the observational side, most of the efforts in studies of ETGs have been focused on improving our current knowledge of the luminous component of these objects, namely the stellar populations and their cosmic evolution \citep[e.g.][]{Fon++04,Cim++06,Poz++10,Pen++10,Cho++14}, while very little is known about the dark matter component.
The underlying dark matter distribution is affected by baryonic physics processes: adiabatic contraction of gas can lead to more concentrated dark matter halos \citep{Gne++11} whereas supernova feedback can remove dark matter from the center of a galaxy \citep{P+G12}.
Observational constraints of dark matter halos can be used to test some of the many models for the effects of baryonic physics on the evolution of ETGs.
Current observational constraints on the dark matter halos of ETGs are scarce and come mostly from the analysis of kinematical tracers data, either alone \citep[see, e.g.,][for recent results]{Cap++13,Agn++14} or in combination with strong gravitational lensing \citep[see, e.g.,][for recent results]{Son++12,New++13,Bar++13,Suy++14}.
The main advantage of strong lensing is that it allows for accurate and precise measurements of masses out to cosmological distances, making it possible to explore the time dimension and address evolutionary questions \citep[see, e.g.,][for a recent review]{Tre10}. 

In this work, we use strong lensing and stellar velocity dispersion measurements for a set of $\sim 80$ lenses from the Strong Lensing Legacy Survey (SL2S) and the Sloan ACS Lens Survey (SLACS) to infer the properties of the population of massive galaxies out to redshift $\sim0.8$.
Using the same sample of lenses, \citet[hereafter Paper IV]{PaperIV} measured the mean density slope $\gamma'$ of the total density profile $\rho(r) \propto r^{-\gamma'}$ across the population of massive ETGs, finding that ETGs evolve while keeping approximately a constant density slope ($d\gamma'/dz = -0.1\pm0.1$).
Although intriguing, a trend of the parameter $\gamma'$ is not of easy
interpretation. 
It is not clear how dark matter and baryons conspire to mantain a constant density slope while the stellar component becomes less concentrated.
Here we address this question by fitting a two-component model, consisting of a stellar spheroid and a dark matter halo, to the same data.
Since dark matter is by definition mass that is not associated with the baryonic component of a galaxy, in order to measure dark matter masses it is necessary to carefully account for all of the mass in stars.
Stellar and dark matter can be effectively disentangled only in systems with data of exceptional quality;
for typical strong lenses, dark matter halo properties can be inferred either by making assumptions about the stellar initial mass function (IMF) \citep{Aug++10}, or by statistically combining information from many systems \citep{R+K05,J+K07,ORF14}, and 
despite recent progress \citep{Cap++12,CvD12,Spi++14}, the true IMF of ETGs is today still a subject of debate \citep{S+L13}.
In this work we study an ensemble of massive ETGs with the goal of 
characterizing simultaneously their distribution of dark matter halo and stellar IMF properties.
We achieve it with a hierarchical Bayesian inference method: a robust statistical tool that allows us to properly take into account scatter in the population. 
We {\it explicitly} take into account the selection function of our lensing surveys, allowing us to learn about the general population of galaxies rather than just characterizing the population of strong lenses.

This paper is organized as follows.  
In \Sref{sect:sample} we describe the sample of lenses used in our study.
In \Sref{sect:twocomp} we describe the model adopted to describe the density profile of the lenses in our sample.
In \Sref{sect:hierarch} we introduce the statistical framework used for the analysis of the population of ETGs.
In \Sref{sect:selfunc} we explain how the selection function of lensing surveys is taken into account.
In \Sref{sect:nfw} we assume a Navarro Frenk and White \citep[NFW][]{NFW97} model for the dark matter halo of all lenses and combine individual measurements to infer the properties of the population of massive ETGs.
In \Sref{sect:gnfw} we generalize the analysis to the case of halos with free inner slope.
After a discussion of our results in
\Sref{sect:discuss} we conclude in \Sref{sect:concl}.
Throughout this paper magnitudes are given in the AB system.  We
assume a concordance cosmology with matter and dark energy density
$\Omega_m=0.3$, $\Omega_{\Lambda}=0.7$, and Hubble constant $H_0$=70
km s$^{-1} $Mpc$^{-1}$.

\section{The sample}\label{sect:sample}
\begin{figure*}
\includegraphics[width=\textwidth]{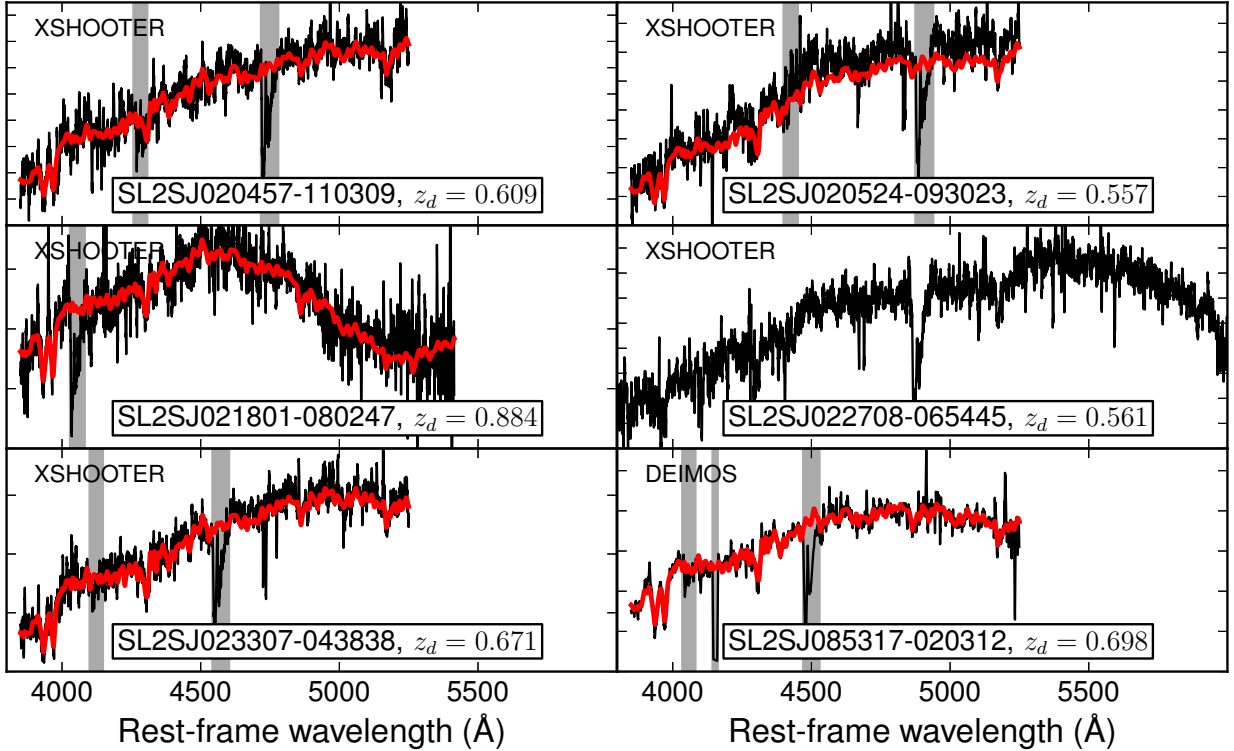}
\caption{1d spectra of new SL2S lenses and lens candidates (in black). Where available, we overplot the best fit spectrum obtained for the velocity dispersion fitting (in red). Only the rest-frame wavelength region used in the fit is shown. Vertical gray bands are regions of the spectrum masked out of the fit and typically correspond to atmospheric absorption features. Each plot indicates the redshift of the galaxy and the instrument used to acquire the data shown.
\label{fig:1dspec}
}
\end{figure*}

\begin{figure}
\includegraphics[width=\columnwidth]{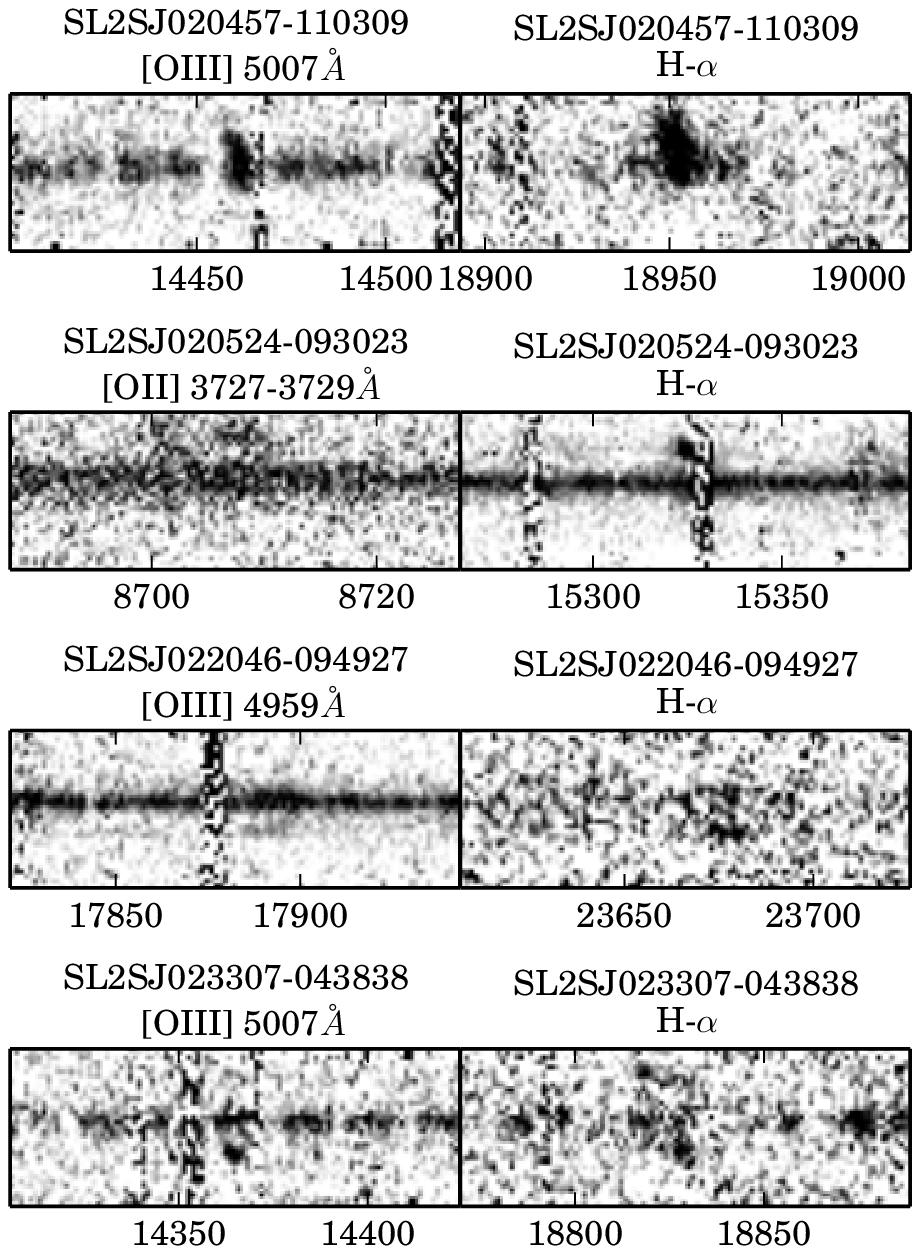}
\caption{2d spectra of new SL2S lenses and lens candidates around detected emission lines from the lensed background source. Observer frame wavelength (in \AA) is labeled on the horizontal axis.
\label{fig:newspec}
}
\end{figure}

Similarly to our previous work in Paper IV, we would like to study the mass distribution of a large sample of galaxies through strong lensing and stellar kinematics and explore dependences of the mass structure on size, stellar mass and redshift.
In order to achieve this goal we need a sample of strong lenses with measurements of the lens and source redshifts, central stellar velocity dispersion and stellar population synthesis (SPS) stellar mass, over a significant range of redshifts.
Similarly to Paper IV, we include in our analysis 25 lenses from the SL2S survey and 53 lenses from the SLACS survey \citep{Aug++10}.
Lens models and SPS stellar masses of the 25 SL2S systems are taken from \citet{PaperIII} (hereafter Paper III), while redshifts and velocity dispersions measurements are reported in Paper IV.
With the intent of increasing the size of the sample of SL2S lenses, we collected new spectroscopic data for eight lenses and lens candidates with the instrument X-Shooter on the Very Large Telescope (P.I. Gavazzi, program 092.B-0663) and with DEIMOS on the W.M. Keck telescope.
These new spectroscopic observations are summarized in \Tref{tab:newspec}.
Three of the objects targeted in these observations had already been observed (see Paper IV). These are systems for which the redshift of either the background lensed source or the main deflector (in the case of SL2SJ021801-080247) was previously unknown. We observed them again with X-Shooter which, thanks to its extended wavelength coverage, increases greatly the chances of detecting emission lines from the lensed sources, as we demonstrated in Paper IV.
For spectra with a sufficiently high signal-to-noise ratio we measured the velocity dispersion of the lens galaxy, necessary for the joint lensing and stellar dynamics analysis carried out in this paper.
Velocity dispersion fits are performed with the same technique described in Paper IV. The measured velocity dispersions are reported in \Tref{tab:newspec}.
One-dimensional spectra of newly observed lenses are plotted in \Fref{fig:1dspec}, and 2d spectra around detected emission lines are shown in \Fref{fig:newspec}.
\renewcommand{\arraystretch}{1.10} 
\begin{deluxetable*}{lccccccccccc}
\tablewidth{0pt}
\tablecaption{\label{tab:newspec} Spectroscopic observations.}
\tabletypesize{\scriptsize}
\tablehead{
\colhead{Name} & \colhead{obs. date} & \colhead{Instrument} & 
\colhead{slit} & \colhead{width} & \colhead{seeing} & \colhead{exp. time} & \colhead{$z_d$} & \colhead{$z_s$} 
& \colhead{$\sigma$} & \colhead{S/N} & \colhead{FWHM} \\
& & & ($''$) & ($''$) & ($''$) & (s) & & & (km/s) & (\AA$^{-1}$) & 
(km/s)
}
\startdata
SL2SJ020457$-$110309 & 11-19-2013 & XSHOOTER & 0.9 & 1.60 & 1.6 & 4140 & 0.609 & 1.89 & $250 \pm 30$ & 9 & $40$ \\ 
SL2SJ020524$-$093023 & 12-03-2013 & XSHOOTER & 0.9 & 1.60 & 1.1 & 2760 & 0.557 & 1.33 & $276 \pm 37$ & 7 & $40$ \\ 
SL2SJ021801$-$080247 & 11-18-2013 & XSHOOTER & 0.9 & 1.60 & 1.0 & 2760 & 0.884 & 2.06 & $246 \pm 48$ & 7 & $40$ \\ 
SL2SJ022046$-$094927 & 11-20-2013 & XSHOOTER & 0.9 & 1.60 & 1.0 & 2760 & 0.572 & 2.61 & $\cdots$ & 7 & $40$ \\ 
SL2SJ022648$-$040610 & 11-20-2013 & XSHOOTER & 0.9 & 1.60 & 1.0 & 2760 & 0.766 & $\cdots$ & $\cdots$ & 6 & $40$ \\ 
SL2SJ022708$-$065445 & 11-23-2013 & XSHOOTER & 0.9 & 1.60 & 0.7 & 2760 & 0.561 & $\cdots$ & $\cdots$ & 9 & $40$ \\ 
SL2SJ023307$-$043838 & 11-24-2013 & XSHOOTER & 0.9 & 1.60 & 0.9 & 2760 & 0.671 & 1.87 & $204 \pm 21$ & 9 & $40$ \\ 
SL2SJ085317$-$020312 & 11-01-2013 & DEIMOS & 1.0 & 1.41 & 0.8 & 9000 & 0.698 & $\cdots$ & $213 \pm 20$ & 14 & $160$ \\ 

\enddata
\tablecomments{Summary of spectroscopic observations and derived parameters.}
\end{deluxetable*}
Lens candidates with no previous spectroscopic observations are also lacking published photometric measurements, lens models and stellar masses, all necessary ingredients for the analysis carried out in this paper. 
We present here these pieces of information.
These newly observed targets all have ground-based photometric data from the instrument MEGA-Cam on the Canada-France Hawaii Telescope (CFHT) in $u, g, r, i, z$ bands. Measurements of the photometric properties of the lens galaxies are performed with the same method adopted in Paper III.
This consists of fitting a de Vaucouleurs profile \citep{deV48} to the data in all five bands simultaneously, while masking out portions of the image contaminated with flux from the (blue) lensed background source.
The measured magnitudes of the lens galaxies are then used to fit stellar population synthesis (SPS) models to infer their stellar masses, again following Paper III.
The stellar masses thus derived depend on the assumed form of the stellar initial mass function (IMF). We infer two sets of SPS masses assuming either a Chabrier \citep{Cha03} or a Salpeter \citep{Salpeter1955} IMF. 
In addition to the five systems with no previous spectroscopic observations, we report stellar mass measurements for SL2SJ021801$-$080247 which are only now possible in virtue of the measurement of the redshift of the lens.
The measured photometric quantities, including SPS stellar masses, are reported in \Tref{tab:newphot}.
\renewcommand{\arraystretch}{1.10} 
\begin{deluxetable*}{lcccccccccc}
\tablewidth{0pt}
\tabletypesize{\small}
\tablecaption{Lens photometric parameters.\label{tab:newphot}}
\tabletypesize{\footnotesize}
\tablehead{
\colhead{Name} & \colhead{$R_{\mathrm{eff}}$} & \colhead{$q$} & \colhead{PA} &
\colhead{$u$} & \colhead{$g$} & \colhead{$r$} & \colhead{$i$} & \colhead{$z$}&
\colhead{$\log{M_*^{\mathrm{(Chab)}}}$} & \colhead{$\log{M_*^{\mathrm{(Salp)}}}$} \\
& (arcsec) & & (degrees) & & & & & & $(M_\odot)$ & $(M_\odot)$ 
}
\startdata
SL2SJ020457-110309 & $1.01$ & $0.67$ & $-20.9$ & $22.81$ & $21.93$ & $20.78$ & $19.82$ & $19.27$ & $11.20\pm0.15$ & $11.46\pm0.15$\\ 
SL2SJ020524-093023 & $0.75$ & $0.64$ & $-75.7$ & $23.69$ & $22.01$ & $20.55$ & $19.50$ & $19.06$ & $11.28\pm0.12$ & $11.52\pm0.12$\\ 
SL2SJ021801-080247 & $1.02$ & $1.00$ & $-49.8$ & $23.05$ & $22.07$ & $21.32$ & $20.33$ & $19.64$ & $11.27\pm0.15$ & $11.54\pm0.14$\\ 
SL2SJ022708-065445 & $0.45$ & $0.28$ & $84.9$ & $23.55$ & $22.49$ & $21.18$ & $20.19$ & $19.76$ & $10.93\pm0.14$ & $11.21\pm0.14$\\ 
SL2SJ023307-043838 & $1.31$ & $0.85$ & $45.9$ & $23.44$ & $21.98$ & $20.63$ & $19.41$ & $19.03$ & $11.44\pm0.14$ & $11.71\pm0.13$\\ 
SL2SJ085317-020312 & $0.85$ & $0.61$ & $16.7$ & $24.45$ & $22.81$ & $21.39$ & $20.12$ & $19.67$ & $11.26\pm0.13$ & $11.51\pm0.13$\\ 

\enddata
\tablecomments{Best fit parameters for de Vaucouleurs models of the surface brightness profile of the lens galaxies, as observed in CFHT data, after careful manual masking of the lensed images.  Columns 2--4 correspond to the effective radius ($R_{\rm eff}$), the axis ratio of the elliptical isophotes ($q$), and the position angle measured east of north (PA).  
The typical uncertainties are $30\%$ on the effective radius, $\Delta q\sim0.03$ for the axis ratio, a few degrees for the position angle, $0.3$ for $u$-band magnitudes, $0.2$ for $g$ and $r$-band magnitudes and $0.1$ for magnitudes in the $i$ and $z$ bands.
The last two columns show the stellar mass measured through stellar population synthesis modeling, assuming a Chabrier or a Salpeter IMF.
}
\end{deluxetable*}

\begin{figure*}
   \includegraphics[width=\textwidth]{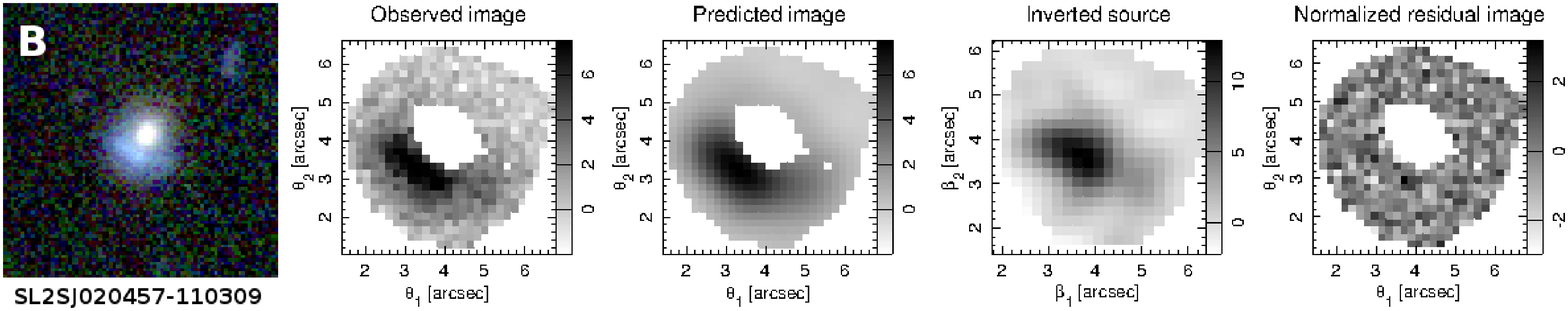} \\
   \includegraphics[width=\textwidth]{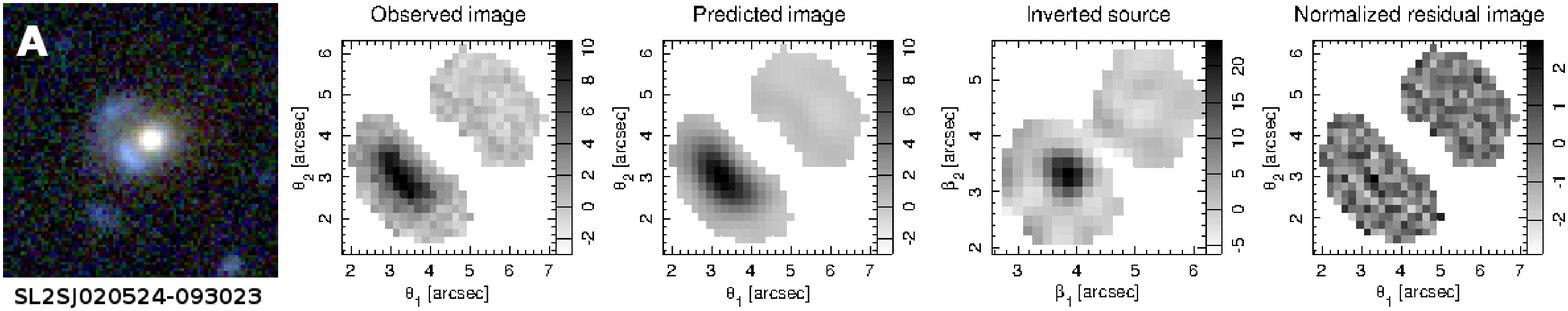} \\
   \includegraphics[width=\textwidth]{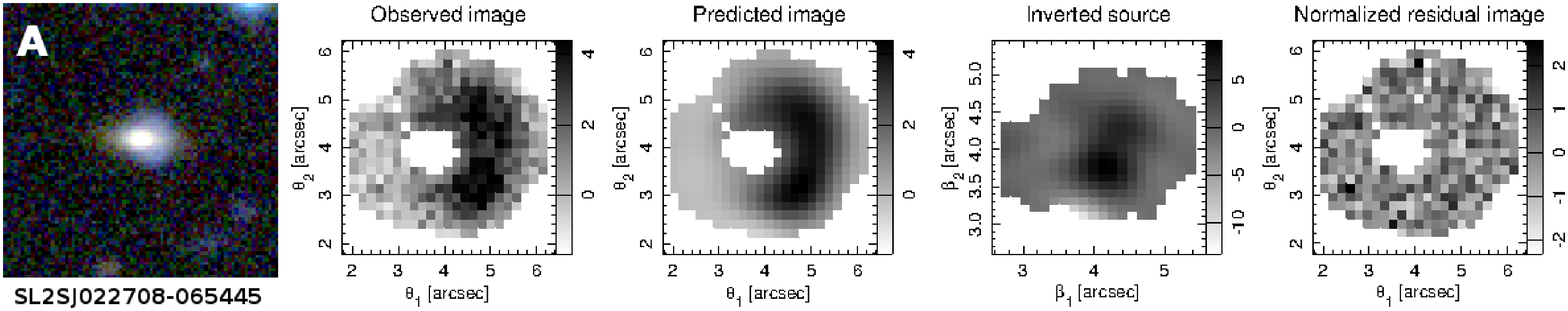} \\
   \includegraphics[width=\textwidth]{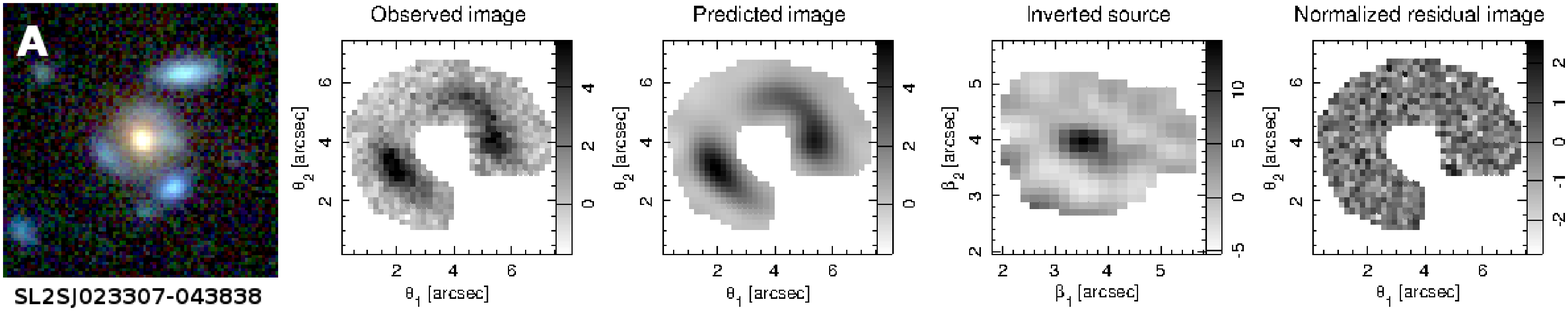} \\
   \includegraphics[width=\textwidth]{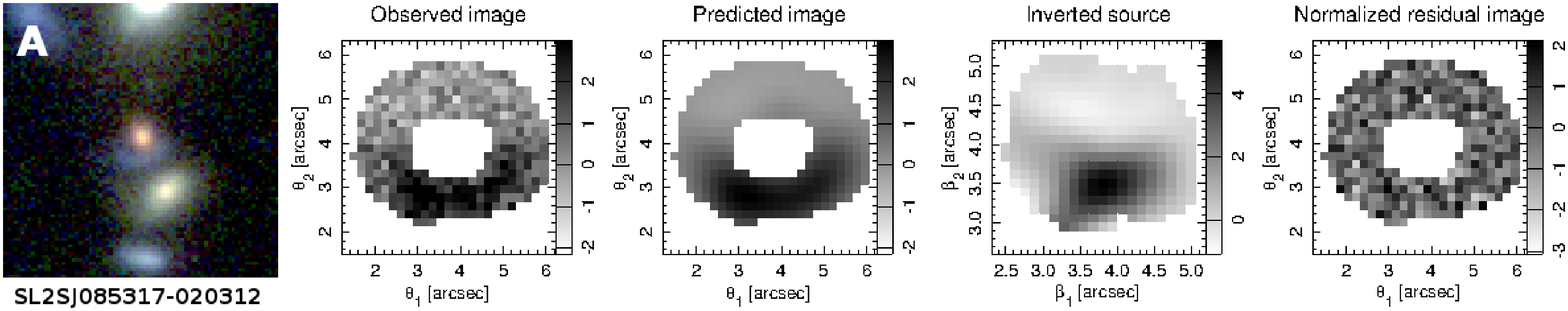} \\
\caption{\label{fig:lensmodels} 
Lens models of five new SL2S galaxy-scale lens candidates.
From left to right: CFHT image, input science image used for the modeling, image reconstruction, lensed source reconstruction, residual image normalized by the uncertainty on each pixel.
}
\end{figure*}

Finally, we present lens models for systems with new spectroscopic observations that had not been analyzed in Paper III.
The lens modeling technique that we adopt here is the same used in Paper III.
We model the mass distribution of each lens as a singular isothermal ellipsoid and fit it by reconstructing the unlensed image of the background source. 
We use the software {\tt GLEE} \citep{S+H10} for this purpose.
Each system is then assigned a grade describing the likelihood of it being a strong lens: grade A for definite lenses, B for probable lenses, C for possible lenses and X for non-lenses.
Images of the lens systems, together with images of the most likely source and image reconstruction are shown in \Fref{fig:lensmodels}.
The inferred lens model parameters and lens grades are reported in \Tref{tab:lensfit}.
We provide here a brief summary of the outcome of the lens modeling of each system.
\begin{itemize}

\item SL2SJ020457-110309. The CFHT data reveal an early-type galaxy with a bright blue image, tangentially elongated with respect to it. The blue object is spectroscopically detected to be at higher redshift than the main galaxy.
The lens model however does not predict the presence of a counter-image. This is probably because the image of the background source appears to be unusually straight, as opposed to the typical arc-like shape of strongly lensed images.
While there is no doubt that the foreground galaxy is lensing the background source, our ground-based data does not allow us to determine whether there is strong lensing or not, therefore we assign a grade B to this system.

\item SL2SJ020524-093023. 
The visible lensed images in this system consist of one arc. The lens model predicts the presence of a counter-image, too faint to be visible in CFHT data.

\item SL2SJ022708-065445. 
An extended blue arc is clearly visible West of a disky early-type galaxy.
The reconstructed source appears to consist of two components close to each other.
In order to achieve a satisfactory fit, we had to put a Gaussian prior on the position angle of the mass distribution, centered on the PA of the light.

\item SL2SJ023307-043838.
This double image system allows us to robustly measure the Einstein radius of the lens galaxy.

\item SL2SJ085317-020312.
One extended arc is visible. We assign a Gaussian prior to the mass position angle in order to obtain a reasonable fit.

\end{itemize}
The systems SL2SJ021801-080247 and SL2SJ022046-094927, which were modeled in Paper III and labeled as grade B systems, are here upgraded to A lenses in virtue of the new spectroscopic data revealing that lens and arc are indeed at two different redshifts.

\renewcommand{\arraystretch}{1.10} 
\begin{deluxetable*}{ccccccc}
\tablewidth{0pt}
\tablecaption{Lens model parameters\label{tab:lensfit}}
\tablehead{
\colhead{Name} & \colhead{$\rein$} & \colhead{$q$} & \colhead{PA} &
\colhead{$m_{\rm{s}}$} & Grade & Notes \\
& (arcsec) & & (degrees) & (mag) & & }
\startdata
SL2SJ020457-110309 & $0.54 \pm 0.07$ & $0.53 \pm 0.19$ & $47.0 \pm 38.4$ & $22.61$ & B &  \\ 
SL2SJ020524-093023 & $0.76 \pm 0.09$ & $0.41 \pm 0.17$ & $-37.0 \pm 23.1$ & $23.74$ & A &  \\ 
SL2SJ022708-065445 & $0.90 \pm 0.05$ & $0.62 \pm 0.10$ & $-85.7 \pm  3.6$ & $24.51$ & A & Disky \\ 
SL2SJ023307-043838 & $1.77 \pm 0.06$ & $0.77 \pm 0.04$ & $50.9 \pm  2.1$ & $24.40$ & A &  \\ 
SL2SJ085317-020312 & $0.93 \pm 0.12$ & $0.52 \pm 0.11$ & $17.6 \pm  5.9$ & $24.52$ & A &  \\ 

\enddata
\tablecomments{Peak value and $68\%$ confidence interval of the posterior
probability distribution of each lens parameter, marginalized over the other
parameters.  Columns 2--4 correspond to the Einstein radius ($\rein$), the axis ratio of the elliptical isodensity contours ($q$), and the position angle measured east of north (PA) of the SIE lens model.  Column 5 shows the magnitude of the de-lensed source in the $g$ filter.
The typical uncertainty on the source magnitude is $\sim0.5$. In Column 6 we report the grade of the lens system, describing the likelihood of it being a strong lens. Column 7 lists notes on the lens morphology.
} 
\end{deluxetable*}

The number of SL2S lenses with a complete set of data necessary for a lensing and dynamics analysis is now 28, with the addition of systems SL2SJ020524-093023, SL2SJ021801-080247 and SL2SJ022046-094927 to the sample analyzed in Paper IV.


\section{Two component mass models}\label{sect:twocomp}

The analysis presented in Paper IV is based on power-law model density profiles for the total (stellar and dark) mass. 
Though very instructive, studying the total density profile leaves open questions on what the detailed structure of the mass profile is.
Different mass profiles could give rise to the same value of $\gamma'$ when fitted with a power-law model.
Massive ETGs have a slope close to $\gamma'\approx 2$.
Models in which the mass follows the light correspond to steeper slopes ($\gamma'\approx2.3$).
In order to get $\gamma'=2$ there must be a non-stellar (dark) component with a mean slope shallower than isothermal.
We want to disentangle the contribution of the dark component to the mass distribution of our lenses from that of the stars.
For this purpose, we consider mass models with two components: a stellar spheroid and a dark matter halo.
We model the stellar spheroid with a de Vaucouleurs profile with effective radius fixed to the observed one, and a uniform prior on the stellar mass-to-light ratio.
The dark matter halo is modeled with a generalized Navarro, Frenk \& White (gNFW) profile \citep{Zha96}:
\begin{equation}\label{eq:gnfw}
\rho_{\mathrm{DM}}(r) \propto \frac{1}{r^{\gammadm}(1 + r/r_s)^{3-\gamma_{\mathrm{DM}}}}.
\end{equation}
Both components are spherical.
We fix the effective radius of the stellar component to the observed one, and the scale radius of the dark matter is fixed to $r_s = 10\reff$, which is a typical value seen in numerical simulations \citep[e.g.][]{Kravtsov13}.
The impact of this choice on our inference will be discussed at the end of this Section.
This mass model has three degrees of freedom, which we describe in terms of the stellar mass $M_*$, the projected dark matter mass within a cylinder of $5\rm{\,kpc}$ radius $\mdm$, and the inner slope of the dark matter halo $\gammadm$.
We fit this model to the observed Einstein radius and central velocity dispersion with the same procedure used in Paper IV. 
Model Einstein radii are calculated given $M_*$, $\mdm$ and $\gammadm$ assuming spherical profiles, and model velocity dispersions are calculated through the spherical Jeans equation assuming isotropic orbits.
The fit is done in a Bayesian framework, assuming a uniform prior on $\log{M_*}$, $\log{\mdm}$ and $\gammadm$, and restricting the range of possible values for the latter quantity to $0.2 < \gammadm < 1.8$. Note that this is very similar to the ``free'' model adopted by \citep{Cap++12}.

\begin{figure}
\includegraphics[width=\columnwidth]{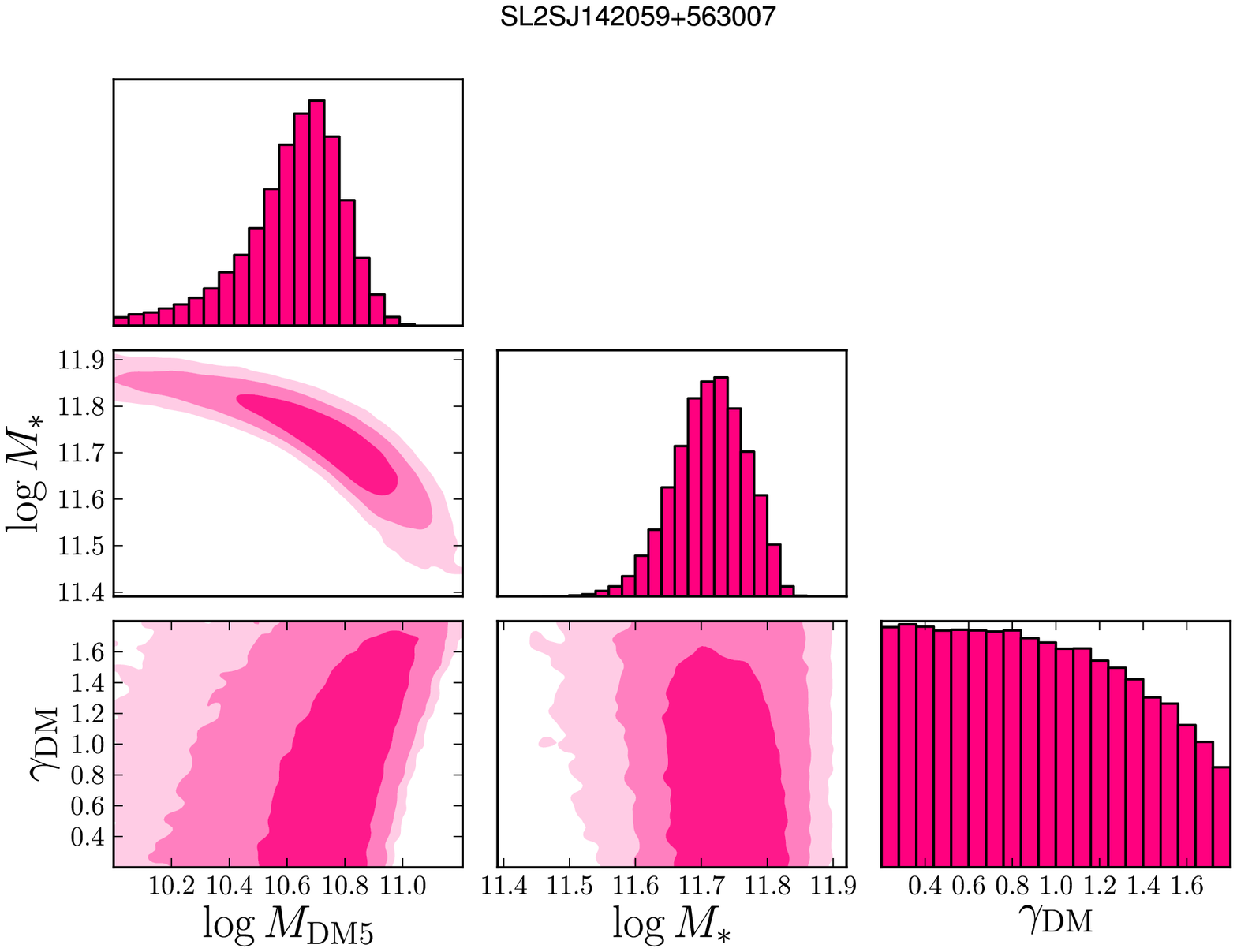}
\caption{\label{fig:example} Posterior probability distribution for a de Vaucouleurs + gNFW mass model of the gravitational lens \example.
The model parameters are the total stellar mass $M_*$, the inner slope of the dark matter halo $\gammadm$ and the projected dark matter mass
$\mdm$
 enclosed within a cylinder of $5$ kpc radius.
}
\end{figure}

As an example, we show in \Fref{fig:example} the posterior probability distribution function of the model parameters for the lens \example.
The model is largely unconstrained, since it consists of three free parameters that are fit to only two pieces of data: the Einstein radius and the central velocity dispersion.
As expected and observed by previous authors \citep[e.g.][]{T+K02a,T+K02b}, there is a strong degeneracy between the inner slope and normalization of the dark matter component.
The tilt of this degeneracy is determined in part by our choice of parametrizing the dark matter halo in terms of the mass enclosed within $5$ kpc. This is not directly observable, while the mass enclosed within the Einstein radius is better constrained by the data. For the lens in this example, the Einstein radius is larger than $5$ kpc, therefore for fixed dark matter mass within $\rein$, the inferred mass at $5$ kpc will depend on the assumed value of the dark matter slope.
Nevertheless, $5$ kpc is close in value to the median Einstein radius of the lenses considered in this work and the choice of $\mdm$ to parametrize the dark matter mass will prove useful later in this work, when analyzing the entire set of lenses statistically.

For systems with data of exceptional quality, the degeneracy between dark matter mass and slope can be broken without having to make additional assumptions \citep[e.g.][]{Son++12}.
In our work, we do not wish to constrain $\gammadm$ and $\mdm$ for individual systems, but we measure their {\it population average} values by statistically combining measurements over a large number of lenses. This will be the subject of Sections \ref{sect:hierarch}, \ref{sect:nfw} and \ref{sect:gnfw}.
Nevertheless, it is interesting to constrain the dark matter content and the stellar mass of individual lenses. 
This can be done, provided we make a more restrictive assumption on the shape of the dark matter halo.
We do this by fixing the inner slope of the dark matter halo to $\gammadm = 1$ and hence restrict ourselves to NFW density profiles for the rest of this section.
The free parameters of the model are now the stellar mass $M_*$ and the normalization of the dark matter halo, $\mdm$.
The model is very similar to the one used by \citet{Tre++10}. The only difference lies in the choice of the scale radius of the NFW component, $r_s$. In \citet{Tre++10} this was fixed to 30 kpc, while here we fix it to ten times the effective radius of the stellar component.

We fit this model to the lensing and stellar kinematics data of each one of the SL2S lenses, as well as lenses from the SLACS survey.
Our two component model, with a de Vaucouleurs spheroid and an NFW dark matter halo, provides excellent fits to most of our lenses. The only exceptions are a few SLACS lenses with very steep density slope $\gamma' > 2.2$, i.e. with relatively large velocity dispersions for their Einstein radius \citep[similar to PG1115+080][]{T+K02b}.
In the context of our model, a steep density slope corresponds to a larger ratio between stellar and dark matter mass, since the NFW halo has a much shallower density profile than a de Vaucouleurs profile at the scale relevant for our measurements, i.e. at the effective radius.
The steepest density profile we can construct with such a two component model is a galaxy with no dark matter. 
For these few SLACS lenses, even if we assign the entire mass enclosed within the Einstein radius to the spheroid, the model velocity dispersion is still smaller than the measured one, although consistent within the uncertainty. 
A perfect match with the data would require $\gammadm>2$, excluded by our prior.
The inference then favors small dark matter masses for those systems.
Adopting a more flexible model for the stellar density profile does not help in this case: \citet{Pos++14} did a similar spheroid and halo decomposition to the same SLACS lenses considered here using a multigaussian fit to the photometry, and still found very small dark matter fractions for some of the objects.

\begin{figure}
\includegraphics[width=\columnwidth]{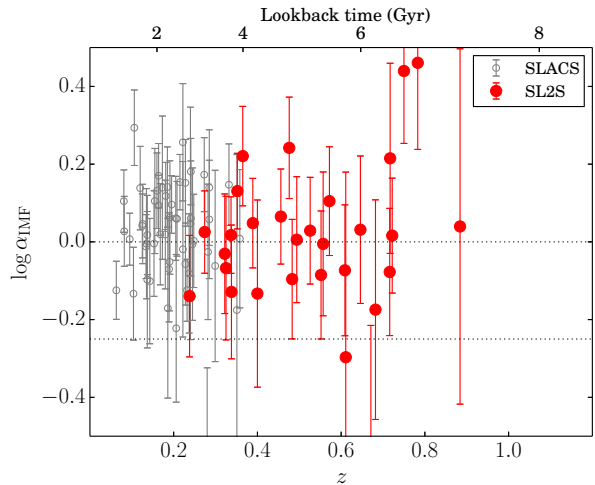}
\caption{\label{fig:alpha_vs_z} IMF mismatch parameter $\alpha_{\mathrm{IMF}} = M_*/M_*^{\mathrm{(Salp)}}$, referred to a Salpeter IMF, as a function of redshift for galaxies of the SL2S, SLACS and LSD samples.}
\end{figure}

\begin{figure}
\includegraphics[width=\columnwidth]{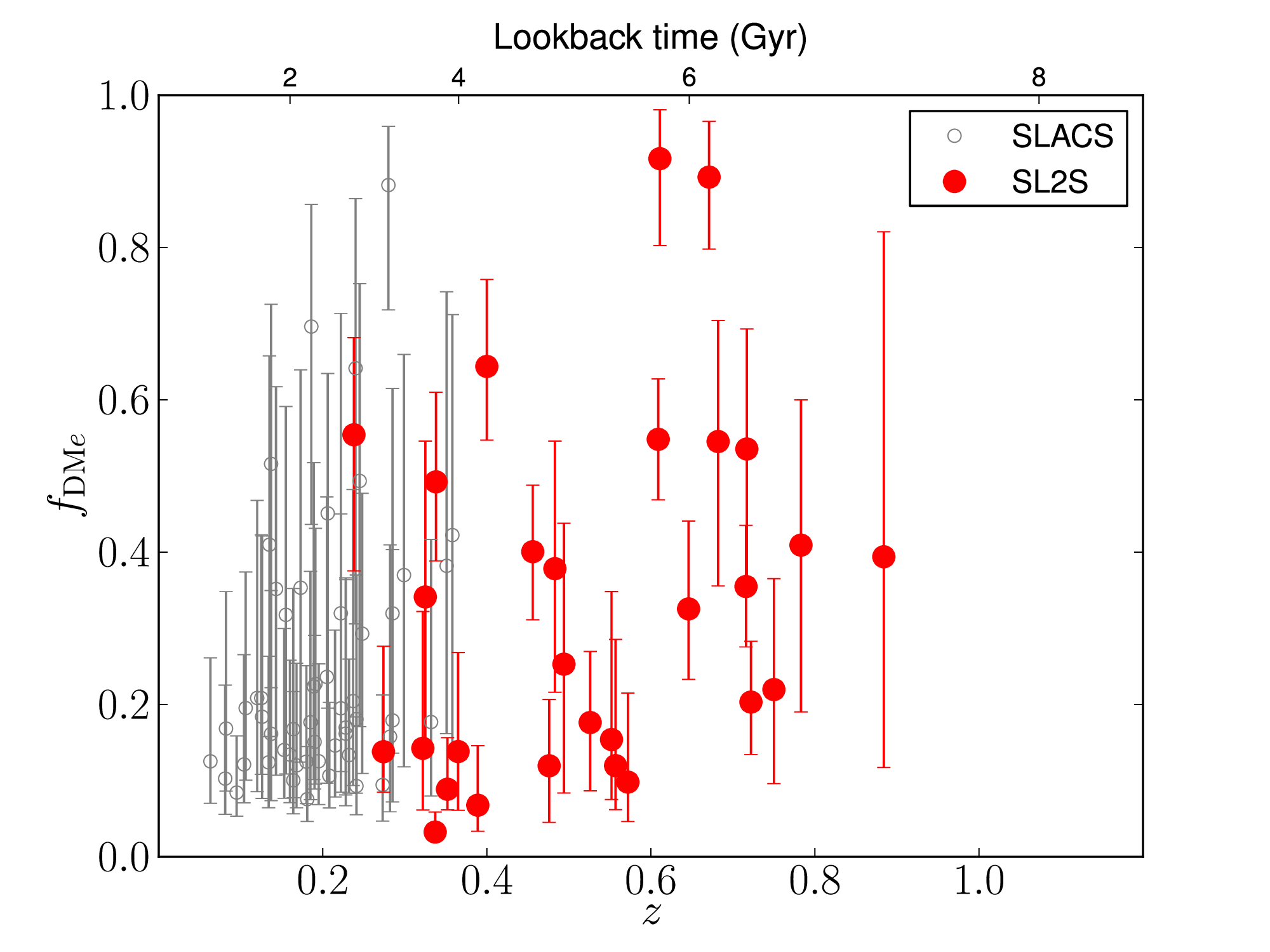}
\caption{\label{fig:fdm_vs_z} Fraction of mass in dark matter projected within a cylinder of radius equal to the effective radius, as a function of redshift.
}
\end{figure}

The derived model parameters for the SL2S lenses are reported in \Tref{table:fitpars}.  The parameters considered are the stellar mass, $M_*$, the projected dark matter mass enclosed within $5$ kpc, $\mdm$, the projected dark matter mass enclosed within $\reff$, $M_{\mathrm{DMe}}$, the fraction of dark matter mass projected within a cylinder of radius $\reff$, $f_{\mathrm{DMe}}$, and finally the {\em IMF mismatch parameter} \citep{Tre++10}, defined as 
the ratio between the true stellar mass and its estimate based on stellar population synthesis models assuming a Salpeter IMF, $\msps$:
\begin{equation}
\aimf \equiv \frac{M_*}{\msps}.
\end{equation}
In this parametrization, a Chabrier IMF corresponds to $\log{\aimf} \approx -0.25$.
Individual measurements of the IMF mismatch parameter and the dark matter fraction are plotted as a function of redshift in Figures \ref{fig:alpha_vs_z} and \ref{fig:fdm_vs_z}.
Under the above assumptions and with typical data quality, we are able to determine dark matter masses with a $\sim50\%$ precision on individual objects.
We recall that the values reported are obtained by assuming a fixed ratio between scale radius of the dark matter halo and effective radius of the light distribution, $r_s = 10\reff$.
Decreasing the value of the proportionality constant to $r_s = 5\reff$ results in dark matter masses smaller by $\sim0.10$ dex and stellar masses larger by $\sim0.05$ dex. We use these values as an estimate of the systematic uncertainty introduced by fixing the value of the dark matter scale radius.
The systematic uncertainty introduced by fixing the dark matter slope is only moderately larger, as can be deduced from \Fref{fig:example}.

Most of the individual measurements of the IMF normalization are consistent with a Salpeter IMF. There are however a few outliers in the measurements shown in \Fref{fig:alpha_vs_z}.
This is because the values of $\aimf$ plotted in \Fref{fig:alpha_vs_z} are obtained by marginalizing over the dark matter mass. 
The actual probability distributions in the $\aimf-\mdm$ space are very elongated and extend closer to the value $\log{\aimf} = 0$ than the marginalized posterior would suggest.
The strong degeneracy between stellar and dark matter mass is taken fully into account in the population analysis described in the next Section.
\renewcommand{\arraystretch}{1.10} 
\begin{deluxetable*}{lcccccccc}
\tablewidth{0pt}
\tabletypesize{\small}
\tablecaption{Stellar and dark matter masses of individual SL2S galaxies, assuming NFW halos.}
\tabletypesize{\footnotesize}
\tablehead{
\colhead{Name} & \colhead{$z$} & \colhead{$\reff$} & \colhead{$\log{M_*^{\mathrm{Salp}}}$} & \colhead{$\log{M_*^{\mathrm{LD}}}$} & \colhead{$\log{\aimf}$} & \colhead{$\log{\mdm}$} & \colhead{$\log{M_{\mathrm{DMe}}}$} & \colhead{$f_{\mathrm{DMe}}$} \\
 & & (kpc) & $M_\odot$ & $M_\odot$ & & $M_\odot$ & $M_\odot$ & 
}

\startdata
SL2SJ020524$-$093023 & $0.56$ & $4.82$ & $11.52\pm0.12$ & $11.54_{-0.11}^{+0.08}$ & $-0.01\pm0.18$ & $10.41_{-0.29}^{+0.37}$ & $10.38_{-0.29}^{+0.37}$ & $0.12_{-0.06}^{+0.17}$ \\ 
SL2SJ021247$-$055552 & $0.75$ & $8.92$ & $11.45\pm0.17$ & $11.90_{-0.09}^{+0.06}$ & $0.44\pm0.19$ & $10.63_{-0.37}^{+0.23}$ & $11.05_{-0.37}^{+0.23}$ & $0.22_{-0.12}^{+0.15}$ \\ 
SL2SJ021325$-$074355 & $0.72$ & $17.67$ & $11.97\pm0.19$ & $12.21_{-0.18}^{+0.12}$ & $0.21\pm0.25$ & $11.04_{-0.17}^{+0.11}$ & $11.96_{-0.17}^{+0.11}$ & $0.54_{-0.17}^{+0.16}$ \\ 
SL2SJ021411$-$040502 & $0.61$ & $6.29$ & $11.60\pm0.14$ & $11.54_{-0.10}^{+0.08}$ & $-0.07\pm0.17$ & $11.16_{-0.06}^{+0.05}$ & $11.32_{-0.06}^{+0.05}$ & $0.55_{-0.08}^{+0.08}$ \\ 
SL2SJ021737$-$051329 & $0.65$ & $4.27$ & $11.53\pm0.16$ & $11.57_{-0.12}^{+0.09}$ & $0.03\pm0.19$ & $11.06_{-0.11}^{+0.09}$ & $10.95_{-0.11}^{+0.09}$ & $0.33_{-0.09}^{+0.12}$ \\ 
SL2SJ021801$-$080247 & $0.88$ & $7.90$ & $11.54\pm0.14$ & $11.75_{-0.53}^{+0.16}$ & $0.04\pm0.46$ & $10.93_{-0.52}^{+0.33}$ & $11.26_{-0.52}^{+0.33}$ & $0.39_{-0.28}^{+0.43}$ \\ 
SL2SJ021902$-$082934 & $0.39$ & $3.01$ & $11.50\pm0.10$ & $11.56_{-0.07}^{+0.04}$ & $0.05\pm0.12$ & $10.48_{-0.30}^{+0.29}$ & $10.13_{-0.30}^{+0.29}$ & $0.07_{-0.03}^{+0.08}$ \\ 
SL2SJ022046$-$094927 & $0.57$ & $3.45$ & $11.36\pm0.11$ & $11.49_{-0.10}^{+0.05}$ & $0.11\pm0.14$ & $10.48_{-0.30}^{+0.30}$ & $10.23_{-0.30}^{+0.30}$ & $0.10_{-0.05}^{+0.12}$ \\ 
SL2SJ022511$-$045433 & $0.24$ & $8.59$ & $11.81\pm0.09$ & $11.67_{-0.13}^{+0.12}$ & $-0.14\pm0.15$ & $11.08_{-0.20}^{+0.11}$ & $11.47_{-0.20}^{+0.11}$ & $0.55_{-0.18}^{+0.13}$ \\ 
SL2SJ022610$-$042011 & $0.49$ & $6.44$ & $11.73\pm0.11$ & $11.76_{-0.13}^{+0.09}$ & $0.01\pm0.16$ & $10.82_{-0.48}^{+0.23}$ & $10.99_{-0.48}^{+0.23}$ & $0.25_{-0.17}^{+0.18}$ \\ 
SL2SJ023251$-$040823 & $0.35$ & $4.78$ & $11.36\pm0.09$ & $11.50_{-0.04}^{+0.03}$ & $0.13\pm0.10$ & $10.22_{-0.15}^{+0.25}$ & $10.19_{-0.15}^{+0.25}$ & $0.09_{-0.03}^{+0.07}$ \\ 
SL2SJ023307$-$043838 & $0.67$ & $9.21$ & $11.71\pm0.13$ & $11.17_{-0.51}^{+0.28}$ & $-0.63\pm0.41$ & $11.35_{-0.04}^{+0.02}$ & $11.79_{-0.04}^{+0.02}$ & $0.89_{-0.09}^{+0.07}$ \\ 
SL2SJ084847$-$035103 & $0.68$ & $3.21$ & $11.24\pm0.16$ & $11.09_{-0.23}^{+0.19}$ & $-0.18\pm0.28$ & $11.18_{-0.15}^{+0.09}$ & $10.87_{-0.15}^{+0.09}$ & $0.55_{-0.19}^{+0.16}$ \\ 
SL2SJ084909$-$041226 & $0.72$ & $3.55$ & $11.63\pm0.13$ & $11.65_{-0.08}^{+0.07}$ & $0.02\pm0.15$ & $11.00_{-0.16}^{+0.11}$ & $10.76_{-0.16}^{+0.11}$ & $0.20_{-0.07}^{+0.08}$ \\ 
SL2SJ084959$-$025142 & $0.27$ & $6.11$ & $11.52\pm0.09$ & $11.56_{-0.06}^{+0.03}$ & $0.03\pm0.10$ & $10.33_{-0.22}^{+0.31}$ & $10.47_{-0.22}^{+0.31}$ & $0.14_{-0.05}^{+0.14}$ \\ 
SL2SJ085019$-$034710 & $0.34$ & $1.35$ & $11.14\pm0.09$ & $11.16_{-0.04}^{+0.03}$ & $0.02\pm0.10$ & $10.25_{-0.18}^{+0.23}$ & $ 9.39_{-0.18}^{+0.23}$ & $0.03_{-0.01}^{+0.03}$ \\ 
SL2SJ085540$-$014730 & $0.36$ & $3.48$ & $11.11\pm0.10$ & $11.35_{-0.10}^{+0.05}$ & $0.22\pm0.13$ & $10.50_{-0.34}^{+0.27}$ & $10.25_{-0.34}^{+0.27}$ & $0.14_{-0.08}^{+0.13}$ \\ 
SL2SJ090407$-$005952 & $0.61$ & $16.81$ & $11.55\pm0.12$ & $11.35_{-0.61}^{+0.33}$ & $-0.30\pm0.48$ & $11.21_{-0.09}^{+0.05}$ & $12.09_{-0.09}^{+0.05}$ & $0.92_{-0.11}^{+0.06}$ \\ 
SL2SJ095921$+$020638 & $0.55$ & $3.47$ & $11.28\pm0.11$ & $11.24_{-0.14}^{+0.05}$ & $-0.09\pm0.16$ & $10.45_{-0.31}^{+0.33}$ & $10.20_{-0.31}^{+0.33}$ & $0.15_{-0.08}^{+0.19}$ \\ 
SL2SJ135949$+$553550 & $0.78$ & $13.08$ & $11.41\pm0.15$ & $11.91_{-0.13}^{+0.09}$ & $0.46\pm0.22$ & $10.74_{-0.36}^{+0.21}$ & $11.44_{-0.36}^{+0.21}$ & $0.41_{-0.22}^{+0.19}$ \\ 
SL2SJ140454$+$520024 & $0.46$ & $11.78$ & $12.10\pm0.10$ & $12.17_{-0.07}^{+0.07}$ & $0.07\pm0.12$ & $11.08_{-0.11}^{+0.08}$ & $11.69_{-0.11}^{+0.08}$ & $0.40_{-0.09}^{+0.09}$ \\ 
SL2SJ140546$+$524311 & $0.53$ & $4.58$ & $11.67\pm0.11$ & $11.71_{-0.09}^{+0.08}$ & $0.03\pm0.14$ & $10.80_{-0.27}^{+0.15}$ & $10.74_{-0.27}^{+0.15}$ & $0.18_{-0.09}^{+0.09}$ \\ 
SL2SJ140650$+$522619 & $0.72$ & $4.35$ & $11.60\pm0.15$ & $11.52_{-0.07}^{+0.07}$ & $-0.08\pm0.17$ & $11.06_{-0.10}^{+0.08}$ & $10.96_{-0.10}^{+0.08}$ & $0.35_{-0.08}^{+0.08}$ \\ 
SL2SJ141137$+$565119 & $0.32$ & $3.04$ & $11.28\pm0.09$ & $11.29_{-0.13}^{+0.05}$ & $-0.03\pm0.16$ & $10.56_{-0.35}^{+0.32}$ & $10.21_{-0.35}^{+0.32}$ & $0.14_{-0.08}^{+0.18}$ \\ 
SL2SJ142059$+$563007 & $0.48$ & $7.86$ & $11.76\pm0.10$ & $11.68_{-0.14}^{+0.11}$ & $-0.10\pm0.16$ & $10.85_{-0.25}^{+0.16}$ & $11.17_{-0.25}^{+0.16}$ & $0.38_{-0.16}^{+0.17}$ \\ 
SL2SJ220329$+$020518 & $0.40$ & $3.86$ & $11.26\pm0.10$ & $11.17_{-0.21}^{+0.14}$ & $-0.13\pm0.24$ & $11.30_{-0.04}^{+0.03}$ & $11.12_{-0.04}^{+0.03}$ & $0.64_{-0.10}^{+0.11}$ \\ 
SL2SJ220506$+$014703 & $0.48$ & $3.93$ & $11.51\pm0.10$ & $11.76_{-0.09}^{+0.07}$ & $0.24\pm0.13$ & $10.76_{-0.38}^{+0.19}$ & $10.60_{-0.38}^{+0.19}$ & $0.12_{-0.07}^{+0.09}$ \\ 
SL2SJ221326$-$000946 & $0.34$ & $2.41$ & $10.99\pm0.10$ & $10.88_{-0.15}^{+0.11}$ & $-0.13\pm0.17$ & $11.06_{-0.07}^{+0.06}$ & $10.57_{-0.07}^{+0.06}$ & $0.49_{-0.10}^{+0.12}$ \\ 
SL2SJ222148$+$011542 & $0.33$ & $5.27$ & $11.55\pm0.09$ & $11.51_{-0.18}^{+0.13}$ & $-0.07\pm0.19$ & $10.88_{-0.34}^{+0.19}$ & $10.92_{-0.34}^{+0.19}$ & $0.34_{-0.19}^{+0.20}$ \\ 

\enddata
\tablecomments{\label{table:fitpars}
Redshifts, effective radii, stellar masses from SPS fitting (from Paper III) and from lensing and dynamics, projected dark matter masses within $5$ kpc and within the effective radius, projected dark matter fractions within the effective radius of SL2S lenses.
}
\end{deluxetable*}

\section{Hierarchical Bayesian Inference}\label{sect:hierarch}

As shown above, the lensing and stellar kinematics data available for typical
strong lenses are not sufficient to constrain both the slope and the
normalization of the dark matter halo of individual objects.
An important question is whether the IMF normalization or the dark matter fraction evolve with time within the population of ETGs.
One possible way of addressing this question is performing a linear fit for $\aimf(z)$ and $\fdm(z)$.
However, the analysis of Paper IV revealed that the density slope $\gamma'$ of ETGs is a function of mass and size as well as redshift. This dependency of $\gamma'$ on $M_*$ and $\reff$ will presumably be reflected on $\aimf$ or $\fdm$.
It is then important to take all dependencies into account when addressing the time evolution of these two parameters.

We want to characterize the population of early-type galaxies which our strong lenses are drawn from.
The focus is on the stellar mass, the IMF normalization, the dark matter mass within 5 kpc and the inner dark matter slope.
Our lenses span a range of redshifts, stellar masses and sizes and we are interested to measure whether there are structural variations with these quantities.
In analogy to the work of Paper IV, the strategy we adopt is a hierarchical Bayesian inference method.
We assume that the values of the parameters describing individual galaxies, $\individ$,
are drawn from a parent distribution described by a set of hyper-parameters $\hyperp$ to be determined from the data $\datad$.
From Bayes theorem,
\begin{equation}
\pr(\hyperp|\datad) \propto \pr(\datad|\hyperp)\pr(\hyperp).
\end{equation}
In turn, the probability of observing the data $\datad$ given the population model $\hyperp$ can be written as the following product over the individual objects' marginal distributions:
\begin{equation}\label{eq:integral}
\pr(\datad|\hyperp) = \prod_i^N \int d \individ  \pr(\datai|\individ)\pr(\individ|\hyperp).
\end{equation}
The integrals run over all the parameters of individual galaxies $\individ$.
The first term in the product of \Eref{eq:integral}
is the likelihood function for an individual galaxy's model parameters $\individ$ given the data $\datai$.
The set of model parameters for an individual galaxy is $\individ = (\mstari,\mdmi,\gammadmi,\aimfi,z_i,\reffi)$, where
$\mdm$ is the dark matter mass within the effective radius.
The data consist of the measured Einstein radius $\reini$, effective radius, redshift, velocity dispersion within the aperture used for spectroscopic observations $\sigma_{\mathrm{ap},i}$ and the stellar mass measured with the stellar population synthesis analysis, $\mspsi$.
In order to speed up computations, redshifts and effective radii of all lenses are assumed to be known exactly.
This approximation does not introduce any significant uncertainty since the typical precision on effective radii measurements is 10\% \citep{PaperIII}, which corresponds to a small uncertainty on the key model parameters, and the uncertainty on redshifts is $\delta z = 0.001$. 
The likelihood of observing data $\datai$ for lens $i$, given its parameters $\individ$ can be factorized as follows:
\begin{eqnarray}\label{eq:likelihood}
\pr(\datai|\individ) = & & \pr(\reini|\individ)\pr(\sigma_{\mathrm{ap},i}|\individ) \times \nonumber \\
& & \pr(\mspsi|\individ) \delta(R_{\mathrm{eff},i}^{\mathrm{(obs)}} - \reffi)\delta(z_i^{\mathrm{(obs)}} - z_i)
\end{eqnarray}
This is possible because the observational uncertainties on the measured Einstein radius, velocity dispersion and SPS stellar mass are independent of each other.
For some lenses in the SL2S sample, more than one velocity dispersion measurements is available \citep{PaperIV}.
In those cases, the velocity dispersion term in \Eref{eq:likelihood} becomes a product over the multiple measurements.

The second term in the integrand of (\ref{eq:integral}) is the probability for the galaxy's individual stellar mass and halo mass
given the set of hyper-parameters $\hyperp$. 
The hyper-parameters must describe the population of galaxies from which our strong lenses are drawn.
Similarly to Paper IV, we assume that the structural properties of ETGs, in this case the dark matter mass and the IMF normalization, are a function of redshift, stellar mass and effective radius.
In the formalism of \citet{Kelly07},
\begin{equation}
\indind = \{z_i,\mstari,\reffi\}
\end{equation}
are the independent variables, while 
\begin{equation}
\depind = \{\mdmi,\gammadmi,\aimfi\}
\end{equation}
are the dependent variables.
It is useful to distinguish among the hyper-parameters the ones that describe the distribution in the independent variables, $\indhyperp$, and those describing the distribution of dependent variables, which we label as $\dephyperp$, following the notation of \citet{Kelly07}. 
The quantity $\pr(\individ|\hyperp)$ has then the following form:
\begin{equation}\label{eq:form_hyperp}
\pr(\individ|\hyperp) = \pr(\indind|\indhyperp) \pr(\depind|\indind,\dephyperp),
\end{equation}
where $\individ = \indind \cup \depind$ and $\hyperp = \indhyperp \cup \dephyperp$.
The probability distribution of the independent variables describes how galaxies in our sample are distributed in the $\{z,M_*,\reff\}$ space.
It encodes both information on the distribution of galaxies in the Universe and the way lens candidates are targeted in our lensing surveys, in terms of selections in stellar mass (or similarly, luminosity), redshift and size.
We assume that the distribution in the independent variables can be written as the product of two Gaussians in $\log{M_*}$ and $\log{\reff}$:
\begin{eqnarray}\label{eq:independent}
& & \pr(\indind|\indhyperp) = \nonumber \\
& & \frac{1}{\msig\sqrt{2\pi}}
  \exp{\left[- \frac{(\log{\mstari}
                      - \mmu(\individ))^2}
                    {2\msig^2} \right]}\times \nonumber \\
& & \frac{1}{\rsig\sqrt{2\pi}}
  \exp{\left[- \frac{(\log{\reffi}
                      - \rmu(\individ))^2}
                    {2\rsig^2} \right]}.
\end{eqnarray}
The mean of these Gaussians is assumed to be different for lenses of different surveys: 
\begin{equation}\label{eq:mstar_sl2s}
\mmusl2s = \mzsl2s(z_i - 0.5) + \log{\mpivsl2s},
\end{equation}
\begin{equation}\label{eq:mstar_slacs}
\mmuslacs = \mzslacs(z_i - 0.2) + \log{\mpivslacs},
\end{equation}
\begin{eqnarray}\label{eq:reff_sl2s}
\rmusl2s = & & \rzsl2s(z_i - 0.5) + \rmsl2s(\log{M_*} - 11.5) + \nonumber \\ 
& &\log{\rpivsl2s},
\end{eqnarray}
\begin{eqnarray}
\rmuslacs = & & \rzslacs(z_i - 0.2) + \rmslacs(\log{M_*} - 11.5) \nonumber \\
& & + \log{\rpivslacs}.
\end{eqnarray}
We also assume different values of the dispersion $\sigma_*$, $\sigma_R$ for SL2S and SLACS lenses.
Note that there's no explicit term for the distribution in $z$ in \Eref{eq:independent}. This is equivalent to assuming a uniform distribution in redshift.
The more physically relevant quantity is the second term in \Eref{eq:form_hyperp}, which describes the properties of the dark matter halos and stellar IMF for galaxies of given $z$, $M_*$ and $\reff$.
The goal of this work is to understand the properties of massive galaxies, irrespective of their lens nature.
However, some galaxies are more likely to be strong lenses than others, because the lensing probability depends in part on the density profile \citep{MvK09}.
Moreover, some strong lenses are more easily detectable than others, as discussed for example by \citet{Arn++12,S+L13,Paper0}.
Then, in order to make accurate statements on the evolution on galaxies based on strong lensing measurements, we must take into account these selection effects.
It is important to verify whether the selection of lenses in the SLACS or SL2S surveys introduce a significant bias with respect to the general population of ETGs, and to quantify it.
The term $\pr(\depind|\indind,\dephyperp)$ should then include a term taking into account the probability for a galaxy described by parameters $\depind$ of being a strong lens detected in a survey. We describe such probability with a set of hyper-parameters $\selhyperp$.
The term relative to the dependent variables is then also assumed to be product of Gaussians, multiplied by a selection function term $\mathcal{S}(\depind|\indind,\selhyperp)$:
\begin{eqnarray}\label{eq:dependent}
& & \pr(\depind|\indind,\dephyperp,\selhyperp) = \mathcal{S}(\depind|\indind,\selhyperp) \times \nonumber \\
& &  \frac{1}{\mdmsig\sqrt{2\pi}}
  \exp{\left[- \frac{(\log{\mdmi}
                      - \mdmmu(\indind))^2}
                    {2\mdmsig^2} \right]}\times \nonumber \\
& &  \frac{1}{\gammasig\sqrt{2\pi}}
  \exp{\left[- \frac{(\gammadmi
                      - \gammamu(\indind))^2}
                    {2\gammasig^2} \right]}\times \nonumber \\
& & 
  \frac{1}{\aimfsig\sqrt{2\pi}}
  \exp{\left[- \frac{(\log{\aimfi}
                      - \aimfmu(\indind))^2}
                    {2\aimfsig^2}\right]}\times \nonumber \\
& & \mathcal{S}(\depind|\indind,\selhyperp).
\end{eqnarray}
The term $\mathcal{S}(\depind|\indind,\selhyperp)$, which will be discussed in \Sref{sect:selfunc}, represents the lensing selection function. This term multiplies the intrinsic distribution of galaxy parameters, which we assumed to be described by a product of Gaussians, to give the distribution observed in strong lenses.
Note that a similar decomposition could in principle be written for the distribution in the independent variables, $\pr(\indind|\indhyperp)$. 
In practice, we are interested in recovering the true distribution for the dependent variables only.
The means of the Gaussians in \Eref{eq:dependent} are in general functions of galaxy redshift, stellar mass and effective radius. 
In particular, we expect the dark matter mass to grow with the stellar mass. We also expect the ratio between stellar and dark mass and the dark matter slope to vary with projected stellar mass density, $\Sigma_* = M_*/(2\pi\reff^2)$, as the results of Paper IV highlighted how the density profile of ETGs at fixed redshift depends to first order on $\Sigma_*$, with systems with more compact stellar distributions having steeper density slopes.
We then choose to parametrize the scaling relations of dark matter halo and stellar IMF normalization in terms of $M_*$ and $\Sigma_*$, as follows:
\begin{eqnarray}\label{eq:mdm_mu}
\mdmmu = & & \mdmz(z_i - 0.3) + \mdmm(\log{\mstari} - 11.5) + \nonumber \\
& & \mdms\log{\Sigma_*/\Sigma_0} + \log{\mdmpiv}
\end{eqnarray}
\begin{eqnarray}
\gammamu = & & \gammapiv 
\end{eqnarray}
\begin{eqnarray}
\aimfmu = & &\aimfz(z_i - 0.3) + \aimfm(\log{M_*} - 11.5) + \nonumber \\
& & \aimfs\log{\Sigma_*/\Sigma_0} + \log{\aimfpiv}
\end{eqnarray}
Although it might seem more natural to assume a scaling in $M_*$ and $\reff$, which would isolate the dependence on stellar mass to only one parameter, $M_*$ and $\reff$ are highly correlated because of the observed tight mass-size relation.
As a result, dependences on $M_*$ or $\reff$ are highly interchangeable and it is difficult to isolate the two with our data.
A parameterization in terms of $M_*$ and $\Sigma_*$ mitigates this effect.
For the average dark matter slope we do not allow for any scaling with any independent variable.
This choice is driven by the little information available from our data on the slope for an individual galaxy (see \Fref{fig:example}). Allowing for too much freedom would result in the average slope of the population of galaxies being undetermined. As always, when the likelihood is not very informative, it is important to choose very carefully the model parameters and priors.
To summarize, the set of hyper-parameters describing the distribution of independent variables is
\begin{equation}
\indhyperp = \{\mz,\mpiv,\msig,\rz,\rmstar,\rpiv,\rsig\},
\end{equation}
with each parameter defined separately for the distribution of SL2S and SLACS lenses, while the hyper-parameters describing the dependent variables distribution is
\begin{eqnarray}
\dephyperp = & & \{\mdmz,\mdmm,\mdms,\mdmpiv,\mdmsig,\gammapiv,\gammasig, \nonumber \\
& & \aimfz,\aimfm,\aimfs,\aimfpiv,\aimfsig,\selhyperp\}.
\end{eqnarray}
Finally, we need to specify the form of the selection function correction $\mathcal{S}(\depind|\indind,\selhyperp)$ in \Eref{eq:dependent}. 
The following section is devoted to it.
%
%
%
%
%
%
\section{The selection function}\label{sect:selfunc}

With the term ``selection function'' we define the mapping between the global population of ETGs and the subset of the population sampled by our lens catalogs.
The goal of this section is to characterize this selection function in a both accurate and computationally tractable way.
SL2S and SLACS, from which our lenses are chosen, are different lensing surveys and are in general subject to different selection effects.
Nevertheless, selection effects for the SL2S and SLACS surveys are qualitatively similar, and will be treated within the same framework.

We can identify three main sources of selection.
The first one is the brightness of the lens. Both SLACS and SL2S samples were assembled by following-up massive ETGs, brighter than a threshold.
For the subset of SLACS galaxies we are considering, the lower limit to the brightness was implicitly set by the requirement of the lens galaxy being targeted in the SDSS spectroscopic survey and having sufficient S/N to allow for a velocity dispersion measurement (citation needed).
For SL2S, only ETGs brighter than $21.5$ in $i$-band were followed-up \citep{Paper0}.
While the luminosity function of ETGs is well described by a one or two Schechter functions \citep{Ilb++13}, selection in brightness results in a different luminosity function for strong lenses, with a cut at low luminosities.
Luminosity is not directly parametrized in the model described in \Sref{sect:hierarch}, but it is tightly related to the stellar mass.
This selection effect can then be captured by the parameters describing the distribution in stellar mass in equations (\ref{eq:mstar_sl2s}) and (\ref{eq:mstar_slacs}).

\begin{figure}
\includegraphics[width=\columnwidth]{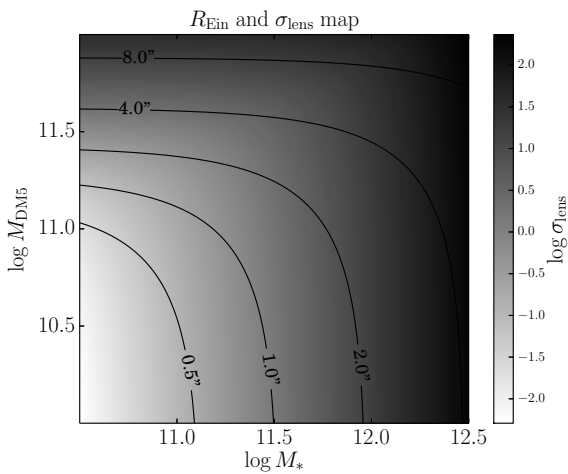}
\caption{\label{fig:sl2s_selfunc} {\em Solid lines:} levels of constant angular Einstein radius (in arcsec) as a function of $\log{M_*}$ and $\log{\mdm}$ for a lens at redshift $z_d = 0.3$, with source redshift $z_s = 1.5$, effective radius $\reff = 5\rm{ kpc}$ and $\gammadm=1$.
{\em Intensity map:} logarithm of the strong lensing cross section, $\scross$ in arcsec$^2$.
}
\end{figure}

The second selection effect is due to different lenses having different strong lensing cross-sections, $\scross$, i.e. different probability of producing systems of multiple images of background sources. 
\citet{MvK09} studied in detail how lensing cross-section depends on lens properties. As expected from general lensing theory, their main finding is that galaxy mass and density profile are the most important parameters determining $\scross$: more massive galaxies have larger lensing cross-section, and so do galaxies with a steeper density profile, at fixed mass.
Quantitatively, the probability of a galaxy described by parameters $\individ$ of being a strong lens is proportional to $\scross(\individ)$. Therefore, the term $\mathcal{S}(\depind|\indind,\selhyperp)$ in \Eref{eq:dependent}, which is proportional to the probability of detecting a lens of parameters $\individ$ given a selection function described by $\boldsymbol{\lambda}$, should also be proportional to $\scross$.
The strong lensing cross-section of a lens with a smooth density profile monotonically decreasing with radius is given by the area enclosed by the radial caustic, i.e. the points in the source plane mapped to points of infinite magnification in the radial direction.
For simplicity, we calculate $\scross(\individ)$ assuming spherical symmetry, as the area enclosed within the radial critical curve, unlike the tangential critical curve, is not very sensitive to the ellipticity of the lens.
Formally, the term $\scross$ has units of solid angle.
In practice, $\scross$ is rescaled so that the probability $\pr(\depind|\indind,\dephyperp,\selhyperp)$ defined in \Eref{eq:dependent} integrated over $\depind$ is normalized to unity.
The third selection effect that we consider is the different detectability of strong lenses of different properties in the two surveys considered, i.e. the probability, given a strong lens, of detecting it in a given lensing survey.
The detection probability in the SL2S was studied by \citet{Paper0}, while that in SLACS-like surveys was studied by \citet{Arn++12}.
The most obvious parameter determining the detection probability is the brightness of the lensed background source: brighter arcs are easier to detect for both the SL2S and SLACS surveys.
In addition to the source brightness, another important parameter determining the detection probability in both SL2S and SLACS is the Einstein radius.
\citet{Paper0} have shown how 
the selection function for SL2S lenses is mostly a function of $\rein$, with a peak in the range $1''<\rein<3''$.
SL2S lenses are selected photometrically by looking for blue arcs around red galaxy in ground based observations \citep{Gav++12}.
This technique works best for lenses with Einstein radius larger than $\sim1''$, since arcs with $\rein$ smaller than that are difficult to resolve in ground based photometry.
The upper limit is due to the fact that lenses with radius smaller than $3''$ were preferentially targeted in the lens-finding algorithm, to favor galaxy-scale lenses over group-scale ones.
For SLACS, lens candidates were selected by looking for emission lines from lensed star-forming galaxies, and then confirmed by HST imaging.
Lenses with too small Einstein radii are more difficult to confirm with this method, because of confusion between the source and the deflector light.
At the opposite end, lenses with too large Einstein radii can escape the selection because the lensed features contribute little to the flux deposited within the $1
\farcs5$-radius fiber used by SDSS spectroscopic observations.
This description matches qualitatively the findings of
\citet{Arn++12}.  The results of \citet{Arn++12} cannot be directly
applied to our analysis though, because the lens models considered by them
have power-law density profiles, different from the
two-component profiles adopted here.  We summarize these properties
instead by approximating the detection probability with a Gaussian
function in $\rein$, which multiplies the previously discussed lensing
cross-section term in the selection correction:
\begin{equation}\label{eq:selfunc}
\mathcal{S}(\depind|\indind,\selhyperp) = \frac{A\scross}{\sqrt{2\pi}\sigma_{\mathrm{sel}}}\exp{\left\{-\frac{(\rein(\individ) - R_{\mathrm{sel}})^2}{2\sigma_{\mathrm{sel}}}\right\}}
\end{equation}
where $\rein$ is a function of the lens parameters $\individ$ and $A$ is a normalization constant.
Here $\selhyperp = \{R_{\mathrm{sel}},\sigma_{\mathrm{sel}}\}$ are hyper-parameters describing the selection function, which can be different for SL2S and SLACS surveys.
Note that there is no source brightness term in \Eref{eq:selfunc}, which we anticipated being important in determining the detection probability of a strong lens. This is because the source brightness is not directly modeled in the hierarchical Bayesian inference framework introduced in the previous section. The term \Eref{eq:selfunc} should then be considered as the effective selection function, obtained by marginalizing over all possible values of the source brightness.
To illustrate what a selection function of the form given by \Eref{eq:selfunc} corresponds to in terms of stellar and dark matter mass, we show in \Fref{fig:sl2s_selfunc} how the Einstein radius of a typical lens changes as a function of $M_*$ and $\mdm$, the other parameters being fixed.
A Gaussian selection function in the Einstein radius implies that only lenses that occupy a band in the $\log{M_*}-\log{\mdm}$ plane can be observed. 
In the same plot we show the lensing cross-section depends on $M_*$ and $\mdm$. As expected, larger masses correspond to larger lensing cross-sections.

\section{Results, NFW halos}\label{sect:nfw}

Before proceeding to analyze the most general case, we focus in this section on models with a fixed dark matter slope $\gammadm = 1$, corresponding to an NFW profile. 
This will indicate whether we can get an adequate description of the evolution of the structure of massive galaxies with a simple dark matter model.
We need to explore the posterior probability distribution $\pr(\hyperp|\bold{d})$ via Markov Chain Monte Carlo (MCMC).
This requires evaluating, for each lens and at each step of the chain, the likelihood term $\pr(\datai|\individ)$ given by \Eref{eq:likelihood} and integrating over all possible values of the lens parameters $\individ$, as given by \Eref{eq:integral}.
The integration over $\log{\aimf}$ can be performed analytically, because both the likelihood \Eref{eq:likelihood} and the parent distribution \Eref{eq:dependent} are Gaussian in $\log{\aimf}$.
Integrals over $z$ and $\log{\reff}$ are trivial, because lens redshift and effective radius are assumed to be known exactly.
We are left with two-dimensional integrals over $\log{M_*}$ and $\log{\mdm}$.
This is a computationally expensive operation, because $M_*$ and $\mdm$ enter the likelihood and the selection function term $\mathcal{S}(\depind|\indind,\selhyperp)$ through the Einstein radius and the velocity dispersion, which are in general non-analytic functions of these parameters. In order to speed up the computation, we sample the likelihood term beforehand for each galaxy and then perform the integrals in \Eref{eq:integral} via Monte Carlo integration at each step of the chain, evaluating the integrand by importance sampling \citep[see e.g.][and references therein]{Suy++10,Bus++11}. 
For both computational and physical reasons (our lenses have a finite amount of stars and dark matter), we truncate the distribution \Eref{eq:dependent} between $10.5$ and $12.5$ in $\log{M_*}$ and between $10.0$ and $12.0$ in $\log{\mdm}$.
In order to be self-consistent, at each step of the chain all probability terms must be normalized to unity. 
The term $\pr(\depind|\indind,\dephyperp)$ requires particular care, as it contains a term, $\mathcal{S}$, that is non-analytic in the model parameters.
The following equality should hold
\begin{equation}
\int d\depind \pr(\depind|\indind,\dephyperp) = 1
\end{equation}
for each set of values of $\indind$ and $\dephyperp$.
This is an implicit equation for the normalization constant in \Eref{eq:dependent}, which we solve via Monte Carlo integration.

\begin{figure*}
\includegraphics[width=\textwidth]{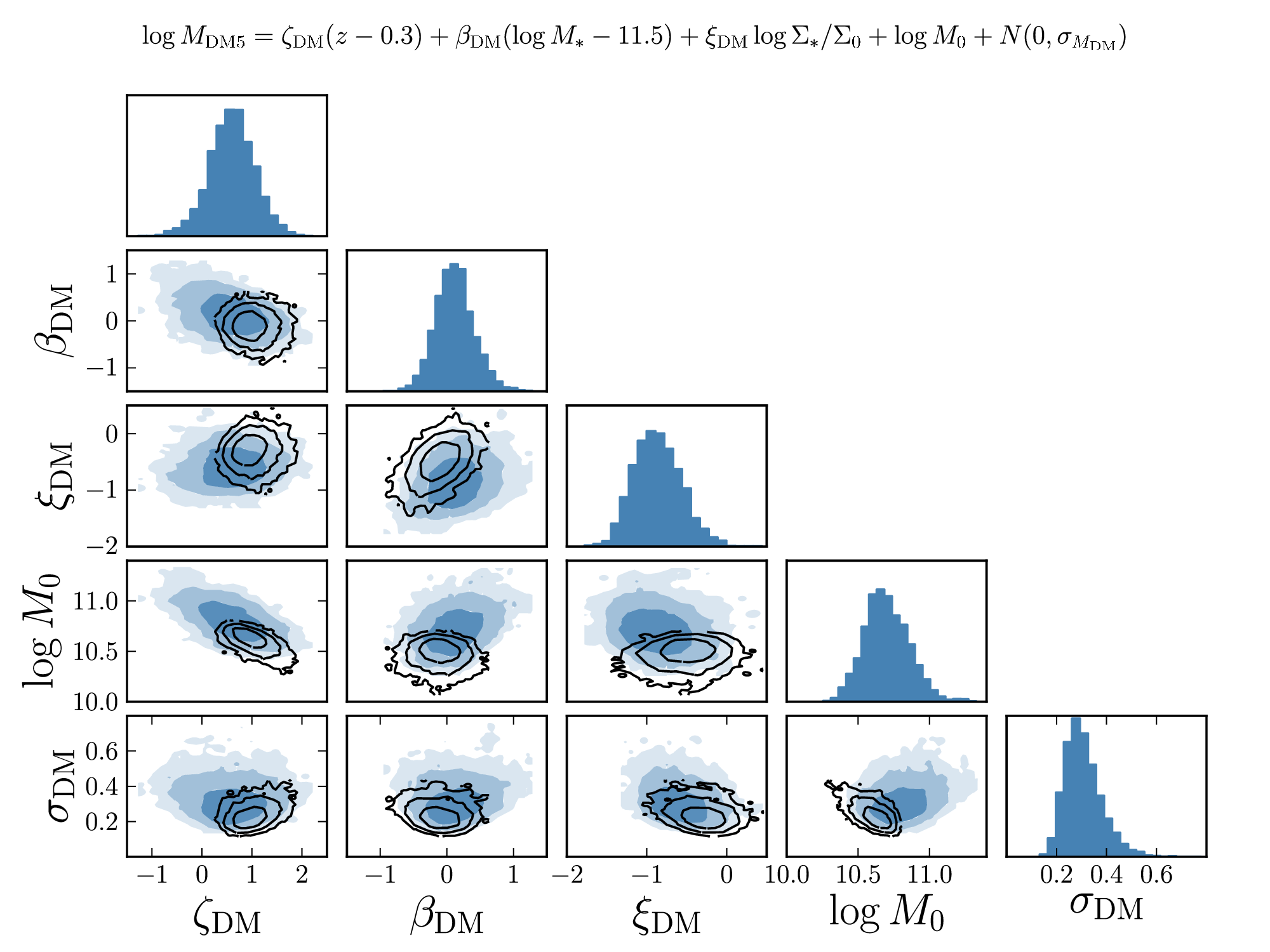}
\caption{\label{fig:mdmpars} Model hyper-parameters describing the dark matter mass within a shell of radius $r_{\mathrm{eff}}$. {\em Empty contours:} inference with no selection function term. {\em Filled contours:} including the selection function term.
The different levels represent the $68\%$, $95\%$ and $99.7\%$ enclosed probability regions. 
}
\end{figure*}

\begin{figure*}
\includegraphics[width=\textwidth]{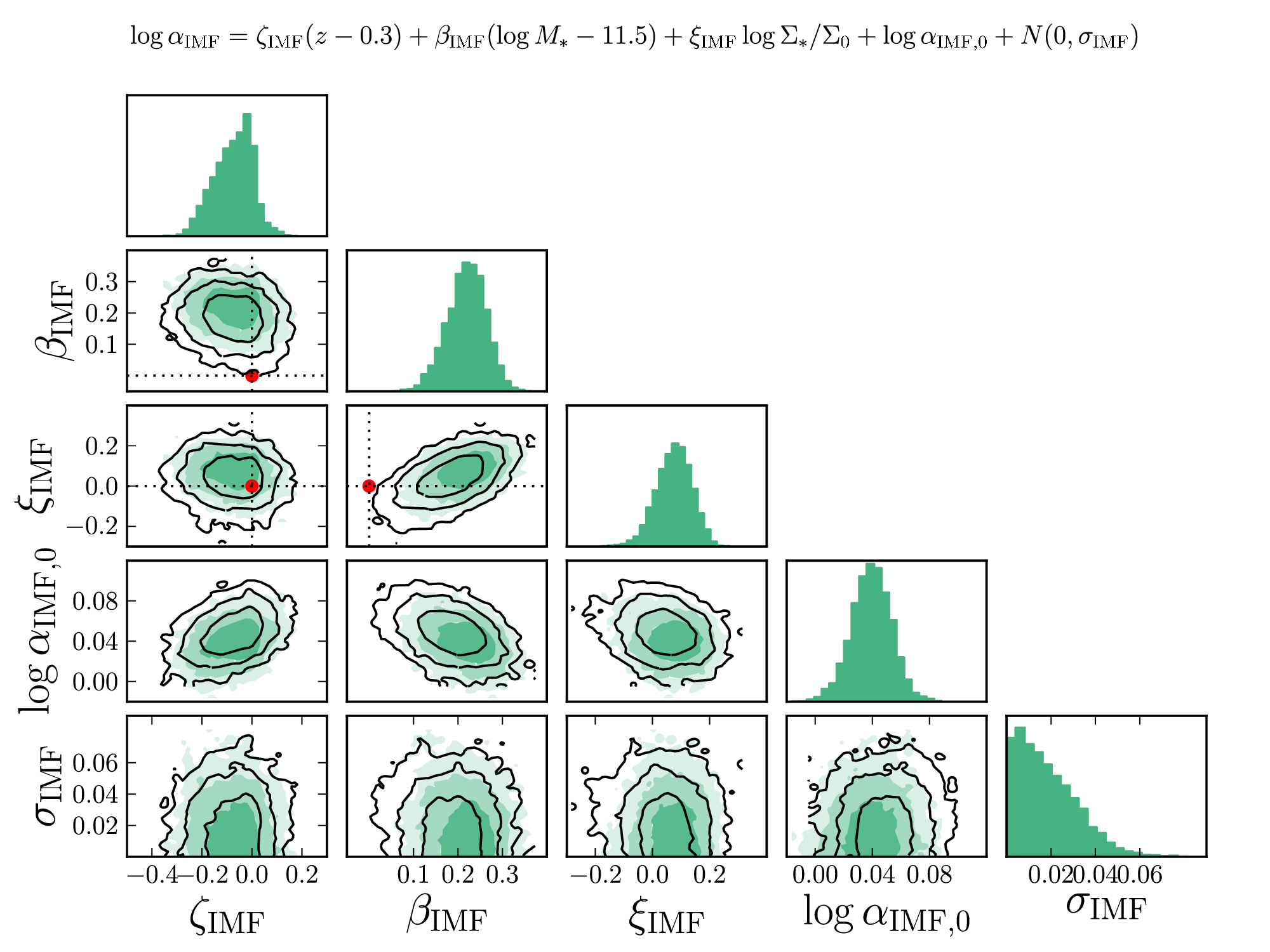}
\caption{\label{fig:imfpars} IMF parameters. The red dots indicate the parameter values corresponding to a universal IMF.
{\em Empty contours:} inference with no selection function term. {\em Filled contours:} including the selection function term.
}
\end{figure*}

We assume a uniform prior on all model hyper-parameters.
We sample the posterior probability distribution with an MCMC with 100000 points, using PyMC \citep{PyMC}.
The median, 16th and 84th percentile of the posterior probability distribution function (PDF) of each parameter, marginalized over the other parameters, are listed in Table~\ref{table:inference}.
The inference on the hyper-parameters describing the dependent variables ($\mdm$ and $\aimf$), $\boldsymbol{\xi}$, is plotted in 
Figures \ref{fig:mdmpars} and \ref{fig:imfpars}.
\renewcommand{\arraystretch}{1.10} 
\begin{deluxetable*}{cccl}
\tablewidth{0pt}
\tabletypesize{\small}
\tablecaption{Bayesian hierarchical inference: the hyper-parameters. NFW model.}
\tabletypesize{\footnotesize}
\tablehead{
 & With $\mathcal{S}$ & No $\mathcal{S}$ & \colhead{Parameter description}
}
\startdata
$\mpivsl2s$ & $11.53_{-0.06}^{+0.06}$ & $11.54_{-0.06}^{+0.06}$ & Mean stellar mass at $z=0.5$, SL2S sample \\ 
$\mzsl2s$ & $0.46_{-0.29}^{+0.32}$ & $0.42_{-0.30}^{+0.31}$ & Linear dependence of mean stellar mass on redshift, SL2S sample \\ 
$\msigsl2s$ & $0.27_{-0.04}^{+0.05}$ & $0.27_{-0.04}^{+0.05}$ & Scatter in mean stellar mass, SL2S sample \\ 
$\mpivslacs$ & $11.66_{-0.04}^{+0.03}$ & $11.67_{-0.03}^{+0.03}$ & Mean stellar mass at $z=0.2$, SLACS sample \\ 
$\mzslacs$ & $2.40_{-0.41}^{+0.39}$ & $2.42_{-0.41}^{+0.36}$ & Linear dependence of mean stellar mass on redshift, SLACS sample \\ 
$\msigslacs$ & $0.24_{-0.02}^{+0.03}$ & $0.23_{-0.02}^{+0.03}$ & Scatter in mean stellar mass, SLACS sample \\ 
$\rpivsl2s$ & $0.69_{-0.03}^{+0.04}$ & $0.68_{-0.03}^{+0.04}$ & Mean effective radius at $z=0.5$, $\log{M_*} = 11.5$, SL2S sample \\ 
$\rzsl2s$ & $0.26_{-0.18}^{+0.21}$ & $0.25_{-0.16}^{+0.22}$ & Linear dependence of mean effective radius on redshift, SL2S sample \\ 
$\rmsl2s$ & $0.65_{-0.14}^{+0.13}$ & $0.70_{-0.13}^{+0.13}$ & Linear dependence of mean effective radius on stellar mass, SL2S sample \\ 
$\rsigsl2s$ & $0.17_{-0.03}^{+0.03}$ & $0.16_{-0.02}^{+0.03}$ & Scatter in mean effective radius, SL2S sample \\ 
$\rpivslacs$ & $0.71_{-0.01}^{+0.01}$ & $0.70_{-0.01}^{+0.01}$ & Mean effective radius at $z=0.2$, $\log{M_*} = 11.5$, SLACS sample \\ 
$\rzslacs$ & $0.09_{-0.18}^{+0.16}$ & $0.02_{-0.19}^{+0.16}$ & Linear dependence of mean effective radius on redshift, SLACS sample \\ 
$\rmslacs$ & $0.61_{-0.05}^{+0.04}$ & $0.65_{-0.04}^{+0.05}$ & Linear dependence of mean effective radius on stellar mass, SLACS sample \\ 
$\rsigslacs$ & $0.07_{-0.01}^{+0.01}$ & $0.07_{-0.01}^{+0.01}$ & Scatter in mean effective radius, SLACS sample \\ 
$\mdmz$ & $0.57_{-0.43}^{+0.44}$ & $0.94_{-0.24}^{+0.25}$ & Linear dependence of $\log{\mdm}$ on redshift. \\ 
$\mdmm$ & $0.10_{-0.24}^{+0.27}$ & $-0.10_{-0.19}^{+0.19}$ & Linear dependence of $\log{\mdm}$ on $\log{M_*}$. \\ 
$\mdms$ & $-0.57_{-0.24}^{+0.27}$ & $-0.27_{-0.19}^{+0.18}$ & Linear dependence of $\log{\mdm}$ on $\log{\Sigma_*}$ \\ 
$\log{\mdmpiv}$ & $10.78_{-0.11}^{+0.14}$ & $10.63_{-0.07}^{+0.06}$ & Mean $\mdm$ at $z=0.3$, $\log{M_*} = 11.5$, $R_{\mathrm{eff}} = 5$ kpc \\ 
$\mdmsig$ & $0.29_{-0.06}^{+0.08}$ & $0.23_{-0.04}^{+0.04}$ & Scatter in the $\mdm$ distribution \\ 
$\aimfz$ & $-0.05_{-0.09}^{+0.06}$ & $-0.06_{-0.09}^{+0.06}$ & Linear dependence of IMF normalization on redshift. \\ 
$\aimfm$ & $0.22_{-0.05}^{+0.04}$ & $0.18_{-0.05}^{+0.05}$ & Linear dependence of IMF normalization on $\log{M_*}$. \\ 
$\aimfs$ & $0.08_{-0.06}^{+0.06}$ & $0.04_{-0.06}^{+0.07}$ & Linear dependence of IMF normalization on $\log{\Sigma_*}$ \\ 
$\log{\aimfpiv}$ & $0.04_{-0.01}^{+0.01}$ & $0.05_{-0.01}^{+0.02}$ & Mean IMF normalization at $z=0.3$, $\log{M_*} = 11.5$, $R_{\mathrm{eff}} = 5$ kpc \\ 
$\aimfsig$ & $0.01_{-0.01}^{+0.02}$ & $0.02_{-0.01}^{+0.02}$ & Scatter in the IMF normalization distribution \\ 
$R_{\mathrm{sel}}^{\mathrm{(SL2S)}}$ & $1.28_{-0.20}^{+0.28}$ & $\cdots$ & Mean observable Einstein radius, SL2S sample \\ 
$\sigma_{\mathrm{sel}}^{\mathrm{(SL2S)}}$ & $0.61_{-0.13}^{+0.18}$ & $\cdots$ & Dispersion in observable Einstein radius, SL2S sample \\ 
$R_{\mathrm{sel}}^{\mathrm{(SLACS)}}$ & $0.95_{-0.24}^{+0.16}$ & $\cdots$ & Mean observable Einstein radius, SLACS sample \\ 
$\sigma_{\mathrm{sel}}^{\mathrm{(SLACS)}}$ & $0.20_{-0.06}^{+0.10}$ & $\cdots$ & Dispersion in observable Einstein radius, SLACS sample \\ 

\enddata
\tablecomments{\label{table:inference}
Median, 16th and 84th percentile of the posterior probability distribution function of each model hyper-parameter, marginalized over the other parameters. Results are reported for the full case and ignoring the selection function.
}
\end{deluxetable*}

The parameters explored by this model are numerous.
Among the results of this analysis we highlight the following.
{\em Under the assumption that dark matter halos of all ETGs have an NFW profile:}
\begin{itemize}
\item The average projected dark matter mass within 5 kpc in massive ETGs is $\log{M_0} = 10.78_{-0.11}^{+0.14}$.
\item We find marginal evidence for an anticorrelation between dark matter mass enclosed within 5 kpc ($\mdm$) and stellar mass density $\Sigma_*$ (parameter $\mdms < 0$), as well as a correlation between $\mdm$ and redshift (parameter $\mdmz < 0$).  
No strong correlations between central dark matter mass and stellar mass is detected (parameter $\mdmm$ is consistent with zero).
\item The IMF normalization is consistent with an IMF slightly heavier than Salpeter: $\log{\aimfpiv} = 0.04\pm0.01$.
\item The IMF normalization correlates positively with stellar mass ($\aimfm = 0.22\pm0.05$). No correlation with redshift or stellar mass density is detected.
\end{itemize}
In order to illustrate the effect of the selection function term $\pr(\individ|\selhyperp)$, we also show the posterior PDF obtained excluding it from our analysis.
Such model without the selection function correction strictly describes the population of strong lenses used in our analysis, as opposed to the general population of massive ETGs.
The posterior PDF of the model without the selection function term is consistent with the more sophisticated model taking into account selection effects.
Nevertheless, the inferred properties of the dark matter halos are slightly different in the two cases.
By not accounting for the selection function we detect a strong dependence of the dark matter mass with redshift, as the parameter $\mdmz$ is larger than zero with more than $3-\sigma$ significance for this model (empty contours in \Fref{fig:mdmpars}).
A positive value of $\mdmz$ means that lenses at lower redshift have preferentially smaller dark matter masses than lenses at higher redshift.
At the same time, the average dark matter mass at the reference point $z=0.3$, $\log{M_*}=11.5$, $\Sigma_* = \Sigma_0$ is smaller with respect to the full analysis: $\log{M_0} = 10.63_{-0.07}^{+0.06}$.
Given the nature of our strong lens sample, with lenses from the SLACS survey dominating the low-redshift part of the sample and SL2S lenses populating the high-redshift end, this result implies that SLACS lenses have on average smaller dark matter masses than similar lenses at higher redshift.
Since the trend in redshift of $\mdm$ is greatly reduced when selection effects are taken into account, this suggests that the lower dark matter masses measured for SLACS lenses is not necessarily related to an intrinsic difference between ETGs at low and intermediate redshift, but might just be the result of the SLACS survey selecting preferentially lenses in smaller dark matter halos.
We further investigated this aspect by repeating the analysis for SLACS and SL2S lenses separately, with and without the selection function term. We confirmed that the SLACS sample is more sensitive to selection effects. In particular it is the Einstein radius selection term of \Eref{eq:selfunc} that drives the offset between the model with $\pr(\individ|\selhyperp)$ and the one without.
According to our model, the detection efficiency in the SLACS survey, which we fit directly for,  is a Gaussian in $\rein$ with mean $R_{\mathrm{sel}} = 0.95_{-0.24}^{+0.16}$ and dispersion $\sigma_{\mathrm{sel}} = 0.20_{-0.06}^{+0.10}$. This is a peaked function in $\rein$ that favors the detection of lenses with smaller Einstein radii and therefore smaller dark matter masses.

We chose to parametrize the dark matter content with the dark matter mass projected within 5 kpc. Many studies, both observational and theoretical, focus instead on the mass enclosed within the effective radius, $\mdme$.
For a better comparison with the literature it is then useful to check what our results imply for this quantity.
As we show in Appendix \ref{app:mdme}, $\mdme$ increases with $M_*$ with a power smaller than unity and has a strong anti-correlation with stellar mass density, meaning that ETGs are not homologous systems.

\subsection{Evolution of individual objects}\label{ssec:nfw_lagrange}

The above analysis reveals how key quantities of massive ETGs scale with redshift, stellar mass and stellar mass density.
In order to gain a better understanding of the evolution of ETGs, it is useful to consider the evolution of individual objects along their evolutionary tracks.
In a fluid kinematics analogy, we would like to transition from an Eulerian description of the fields $\mdm$ and $\aimf$ at fixed $M_*$, $z$ and $\reff$, which is given by the analysis presented above, to a Lagrangian description of the evolution of these quantities along the history of individual galaxies.
While the latter quantity is not directly observable, it can be inferred with the formalism introduced in Paper IV, which connects the observed scaling relations with external constraints on the evolution of size and stellar mass.
We denote with $d/dz$ the derivative with respect to redshift along the evolutionary track of an individual galaxy.
Then we can write
\begin{eqnarray}\label{eq:mdm_lagrange}
\frac{d\log{\mdm}}{dz} = & & \frac{\partial\log{\mdm}}{\partial z} + \frac{\partial \log{\mdm}}{\partial \log{M_*}}\frac{d\log{M_*}}{dz}+ \nonumber \\
& & \frac{\partial \log{\mdm}}{\partial \log{\Sigma_*}}\frac{d\log{\Sigma_*}}{dz}
\end{eqnarray}
and
\begin{eqnarray}\label{eq:imf_lagrange}
\frac{d\log{\aimf}}{dz} = & & \frac{\partial\log{\mdm}}{\partial z} + \frac{\partial \log{\mdm}}{\partial \log{M_*}}\frac{d\log{M_*}}{dz}+ \nonumber \\
& & \frac{\partial \log{\aimf}}{\partial \log{\Sigma_*}}\frac{d\log{\Sigma_*}}{dz}.
\end{eqnarray}
This is the evolution in $\mdm$ and $\aimf$ of a galaxy for which these quantities scale with $z$, $M_*$ and $\Sigma_*$ in the same way as the population averages $\mu_{\mathrm{DM}}$ and $\mu_{\mathrm{IMF}}$.
The presence of scatter will in general modify the picture, but we expect the above expressions to be correct to first order.
Equations \ref{eq:mdm_lagrange} and \ref{eq:imf_lagrange} hold as long as the population of ETGs is not significantly polluted by the formation of new objects that enter the sample in the redshift range that we consider.
Current estimates show that the number density of massive galaxies evolves very modestly below redshift 1 \citep{Cas++13,Ilb++13,Muz++13}.

\begin{figure}
\includegraphics[width=\columnwidth]{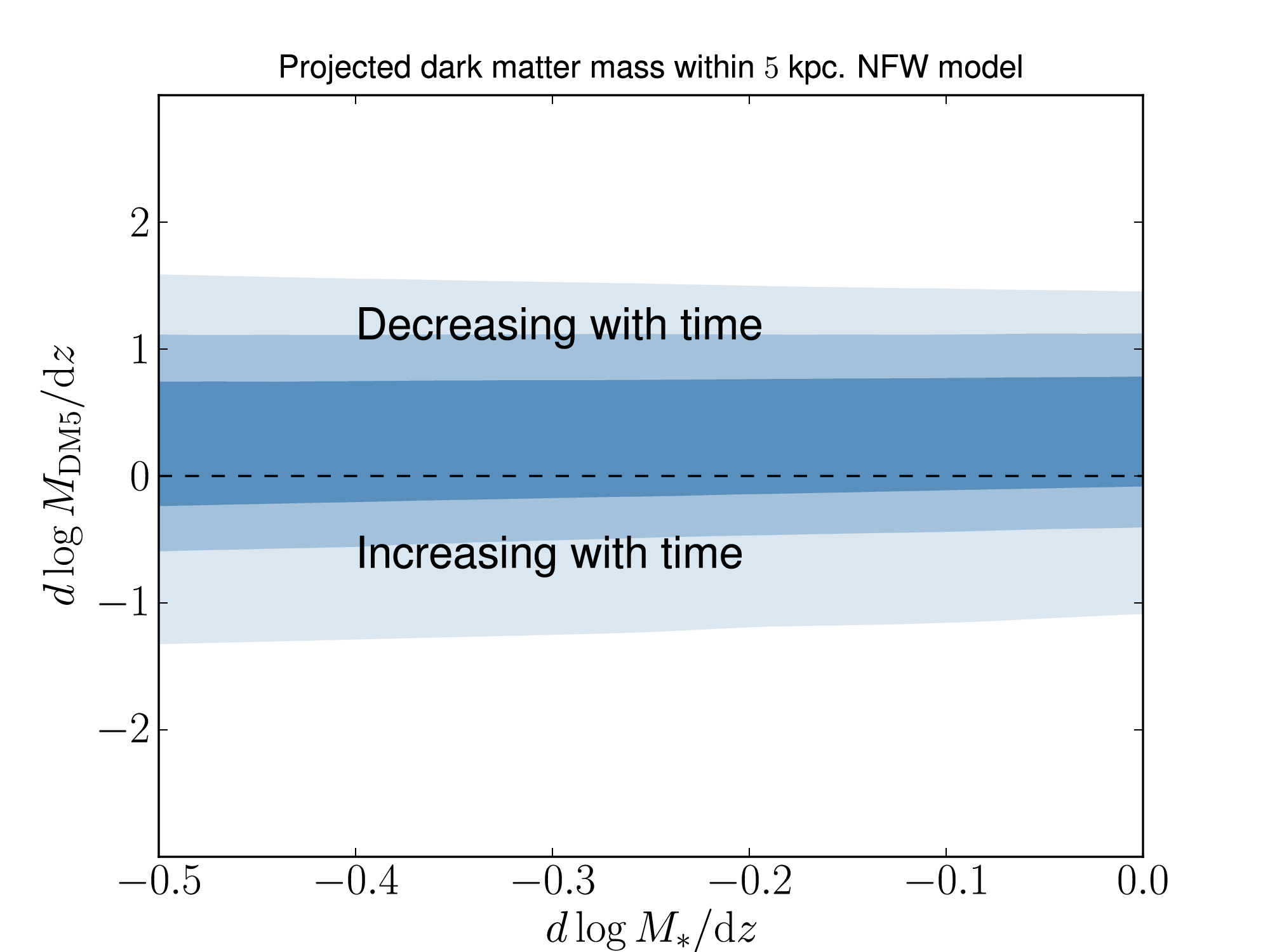}
\caption{\label{fig:mdm_lagrange_nfw} Rate of change in projected dark matter enclosed within a cylinder of radius $5$ kpc along the evolutionary track of an individual galaxy, calculated from \Eref{eq:mdm_lagrange}, as a function of the growth rate in stellar mass. An NFW profile for the dark matter halo is assumed and selection effects are taken into account.
The different colors represent the 68\%, 95\% and 99\% probability regions.
}
\end{figure}

\begin{figure}
\includegraphics[width=\columnwidth]{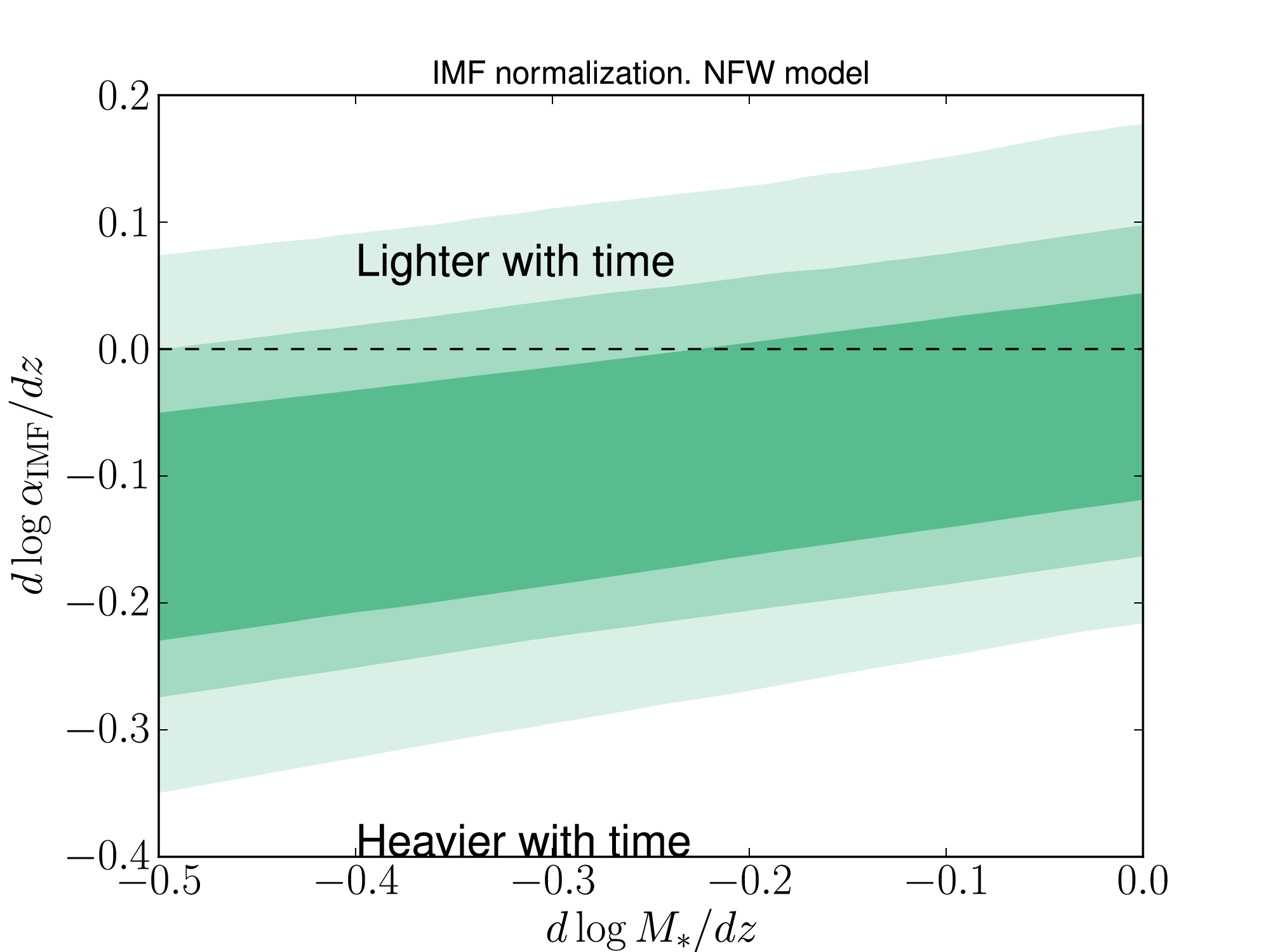}
\caption{\label{fig:imf_lagrange_nfw} Rate of change in the IMF normalization along the evolutionary track of an individual galaxy, calculated from \Eref{eq:imf_lagrange}, as a function of the growth rate in stellar mass. An NFW profile for the dark matter halo is assumed and selection effects are taken into account.
}
\end{figure}

The partial derivatives in \Eref{eq:mdm_lagrange} can be identified with the parameters $\mdmz$, $\mdmm$ and $\mdms$ measured in our analysis, while those in \Eref{eq:imf_lagrange} are matched to $\aimfz$, $\aimfm$ and $\aimfs$.
The two total derivatives, $d\log{M_*}/dz$ and $d\log{\Sigma_*}/dz$
are the rate of change in stellar mass and stellar mass density of an individual galaxy.
The latter depends on the former, and on the evolution of the effective radius as well:
\begin{equation}
\frac{d\log{\Sigma_*}}{dz} = \frac{d\log{M_*}}{dz} - 2\frac{d\log{\reff}}{dz}.
\end{equation}
As in Paper IV, we can evaluated $d\log{\reff}/dz$ by combining constraints from the redshift and mass dependence of $\reff$, assuming again that individual galaxies evolve in the same way as the average:
\begin{equation}
\frac{d\log{\reff}}{dz} = \frac{\partial \log{\reff}}{\partial z} + \frac{\partial\log{\reff}}{\partial \log{M_*}}\frac{d\log{M_*}}{dz}.
\end{equation}
For the scaling of effective radius with mass, we take the value measured by \citet{New++12b}: $\partial \log{\reff}/\partial \log{M_*} = 0.59\pm 0.07$.
The redshift dependence has been measured by a number of authors \citep[e.g.][]{Dam++11,New++12b,Cim++12,Hue++13}, with significant scatter between the measurements. Here we take $\partial \log{\reff}/\partial z$ to be the mean between these measurements, and use the standard deviation as an estimate of its uncertainty: $\partial \log{\reff}/\partial z = -0.37 \pm 0.08$.
With this prescription we evaluate the derivatives \Eref{eq:mdm_lagrange} and \Eref{eq:imf_lagrange}, which we plot in \Fref{fig:mdm_lagrange_nfw} and \Fref{fig:imf_lagrange_nfw} as a function of the, unknown, mass growth rate $d\log{M_*}/dz$.
\begin{figure}
\includegraphics[width=\columnwidth]{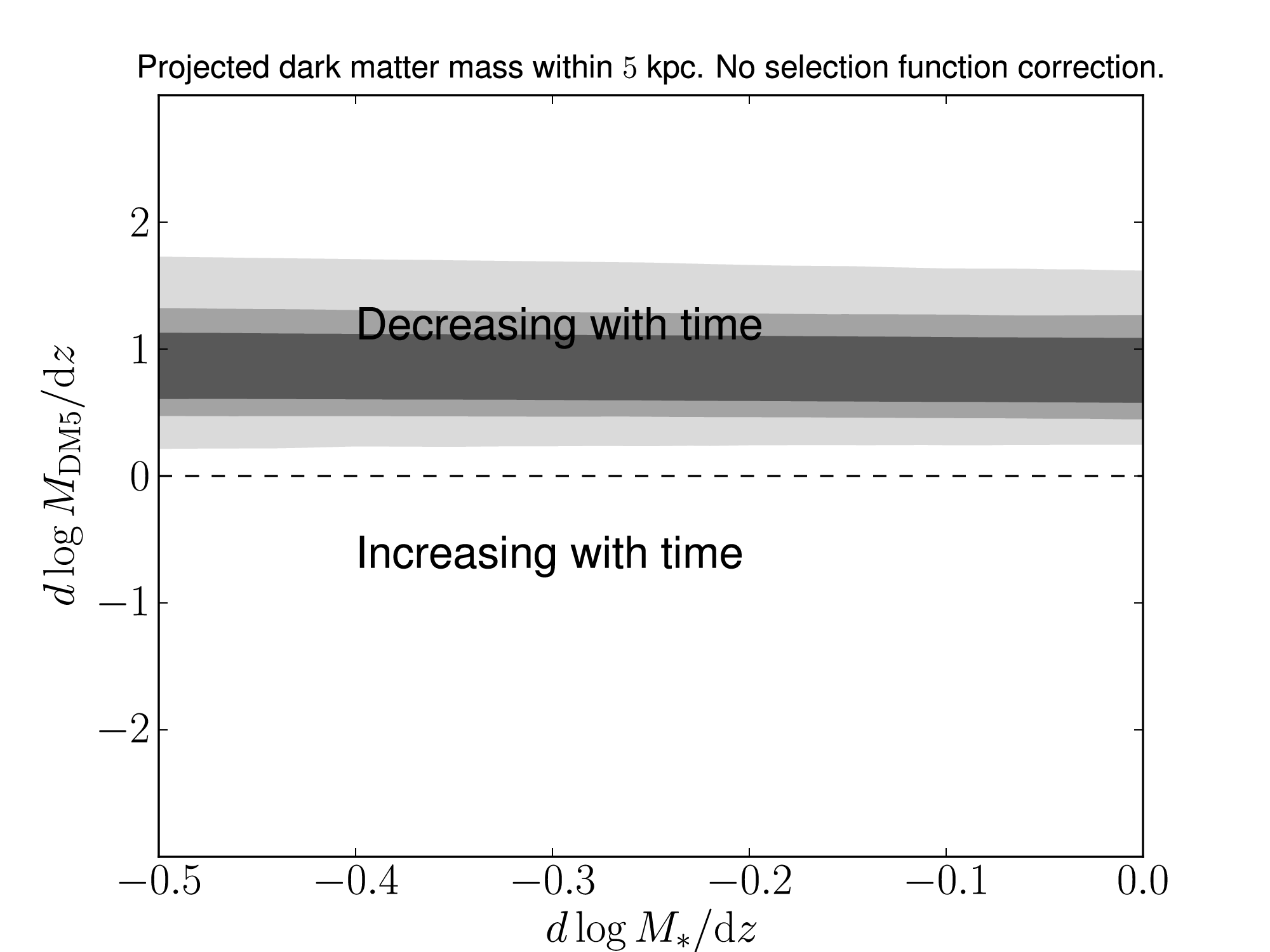}
\caption{\label{fig:mdm_lagrange_nosel_nfw} Rate of change in projected dark matter enclosed within a cylinder of radius $5$ kpc along the evolutionary track of an individual galaxy, calculated from \Eref{eq:mdm_lagrange}, as a function of the growth rate in stellar mass, inferred ignoring the selection function term.
An NFW profile for the dark matter halo is assumed.
The different colors represent the 68\%, 95\% and 99\% probability regions.
}
\end{figure}
The uncertainties on the derived evolution of enclosed dark matter and IMF normalization are relatively large, in part due to the uncertainty on the mass-size relation and its evolution.
We expect the dark matter enclosed within $5$ kpc to exhibit little change over time, since most of the matter accreted in the later phases of the evolution of an ETGs will likely grow the outskirts of the galaxy.
We also expect the IMF normalization to show little change over time, because a significant change of the IMF would require the accretion or formation of stars with an extremely different IMF from the preexisting population, a scenario at odds with our current knowledge of stellar populations in the Universe.
Our measurements are consistent with these expectations, though with the current data we are unable to make precise statements in this regard. 

In \Fref{fig:mdm_lagrange_nosel_nfw} we plot the evolution in dark matter mass inferred from the population model without the selection function term -- that is to say, assuming that the strong lenses from both the SL2S and SLACS survey are an unbiased sample of the general population of early-type galaxies.
Under this assumption, the data require dark matter masses to {\it decrease with time} at a significant rate, with more than 3-$\sigma$ confidence, in sharp contrast with the result plotted in \Fref{fig:mdm_lagrange_nfw}, which does take the selection function into account.
It is difficult to imagine a physical scenario in which the stellar mass increases by a modest amount while at the same time a comparable, or larger, amount of dark matter is ejected from the inner $5$ kpc of a galaxy.
We believe that the implausible scenario of \Fref{fig:mdm_lagrange_nosel_nfw} is an indication that the selection function does indeed need to be included in the modeling. (However, as we will show below, the lack of selection function modeling in our previous work does not actually 
affect the conclusions of papers I-IV.)


\section{Results, free inner slope}\label{sect:gnfw}

The results of the analysis presented in \Sref{sect:nfw} depend on the assumption of a fixed NFW shape for the dark matter profile of all ETGs.
Here we relax that assumption and consider gNFW profiles instead, with density profile given by \Eref{eq:gnfw}.
We impose that individual values of the inner dark matter slope lie in the range $0.2 < \gammadm < 1.8$, as we expect the dark matter density profile to be shallower than the total density profile, which is measured to be close to isothermal \citep[$\gamma'\approx 2$][]{Koo++06}.
As pointed out in \Sref{sect:twocomp}, allowing for one extra degree of freedom in the dark matter halo model results in a significant degeneracy in the determination of the properties of individual galaxies.
However, we know that ETGs constitute a family of objects with rather homogeneous characteristics. The large number of available lenses therefore can help us break the degeneracy and pin down the population average properties of the luminous and dark matter distributions.
In particular, the tilt of the degeneracy contour between the dark matter mass within $5$ kpc and the inner slope, plotted in \Fref{fig:example} for one of the SL2S lenses, depends on the value of the Einstein radius: the data constrain the projected mass enclosed within $\rein$ and the value of $\mdm$ is obtained by extrapolating the Einstein mass to $5$ kpc assuming a value of $\gammadm$.
Different lenses have different values of $\rein$, therefore the direction of the degeneracy contour between $\mdm$ and $\gammadm$ will be different for each lens, depending on the amount of extrapolation required to match the mass at $5$ kpc from the mass at $\rein$.
If the scatter in $\mdm$ across the population of massive ETGs is small, then it is possible to rule out extreme values of the dark matter slope by simply multiplying the probability distribution for individual lenses, which is what our hierarchical Bayesian  model effectively does.

\begin{figure*}
\includegraphics[width=\textwidth]{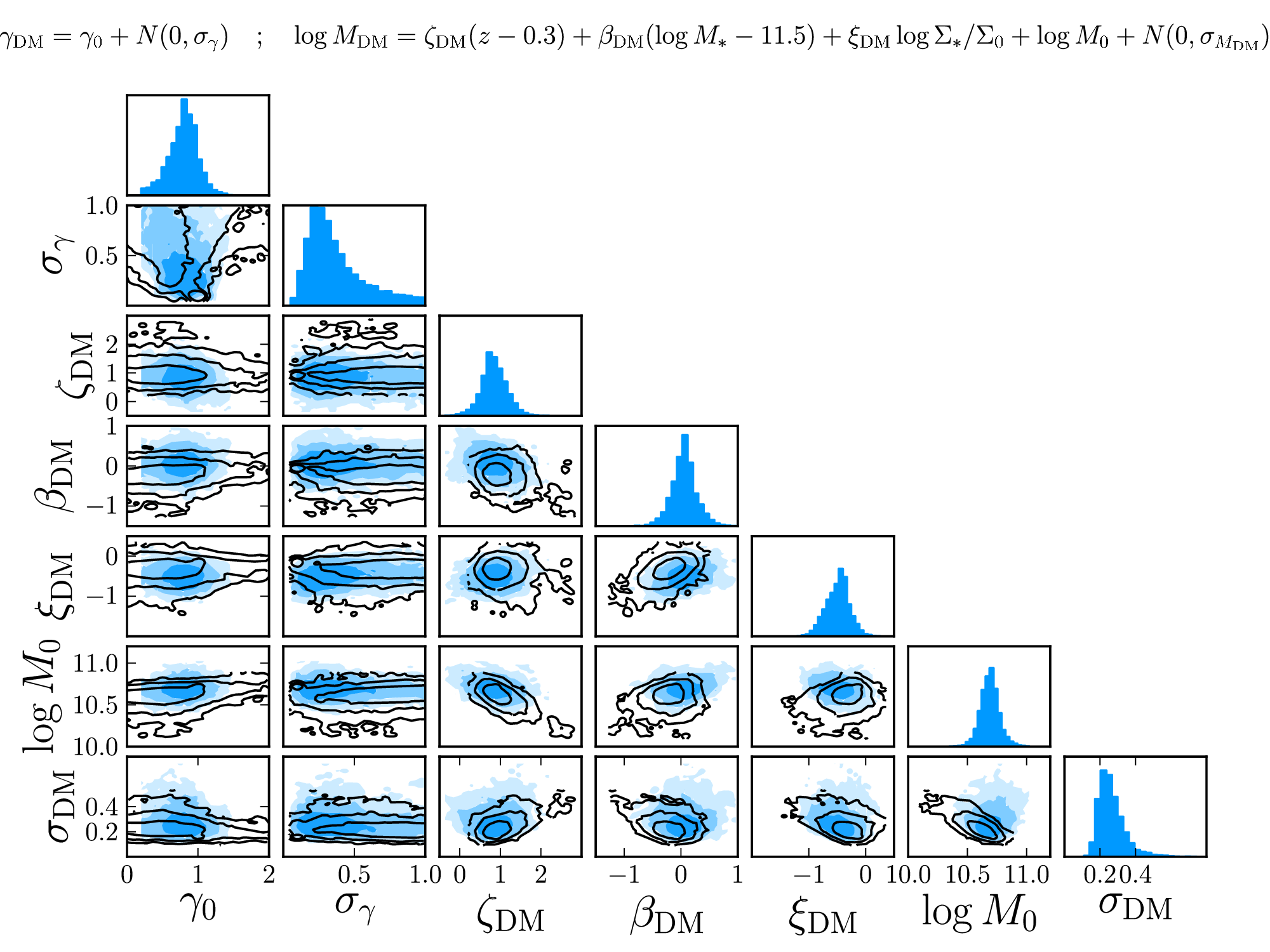}
\caption{\label{fig:gnfw_mdm} Model hyper-parameters describing the dark matter mass within a shell of radius $r_{\mathrm{eff}}$ and inner slope, for a gNFW dark matter halo. 
{\em Empty contours:} inference with no selection function term. {\em Filled contours:} including the selection function term.
The different levels represent the $68\%$, $95\%$ and $99.7\%$ enclosed probability regions.
}
\end{figure*}

\begin{figure*}
\includegraphics[width=\textwidth]{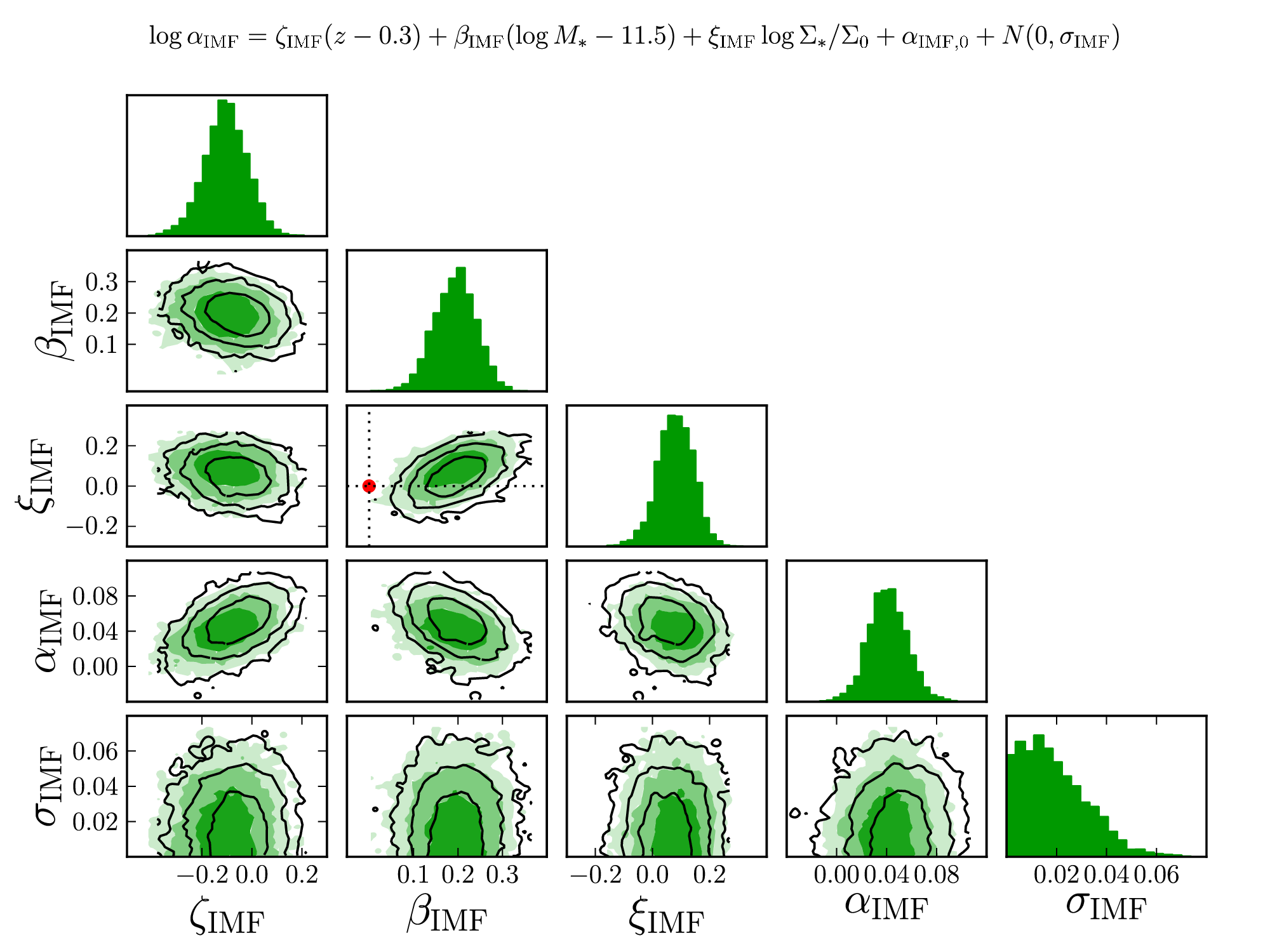}
\caption{\label{fig:gnfw_imf} IMF normalization hyper-parameters, for a gNFW dark matter halo. The red dot indicates the parameter values corresponding to a universal IMF.
{\em Empty contours:} inference with no selection function term. {\em Filled contours:} including the selection function term.
}
\end{figure*}

The posterior PDF for the parameters describing the population distribution of dark matter halos and IMF normalizations is plotted in \Fref{fig:gnfw_mdm} and \Fref{fig:gnfw_imf}, while the median and 68\% confidence interval is listed in \Tref{tab:gnfw_inference} for all the inferred parameters.
\begin{figure}
\includegraphics[width=\columnwidth]{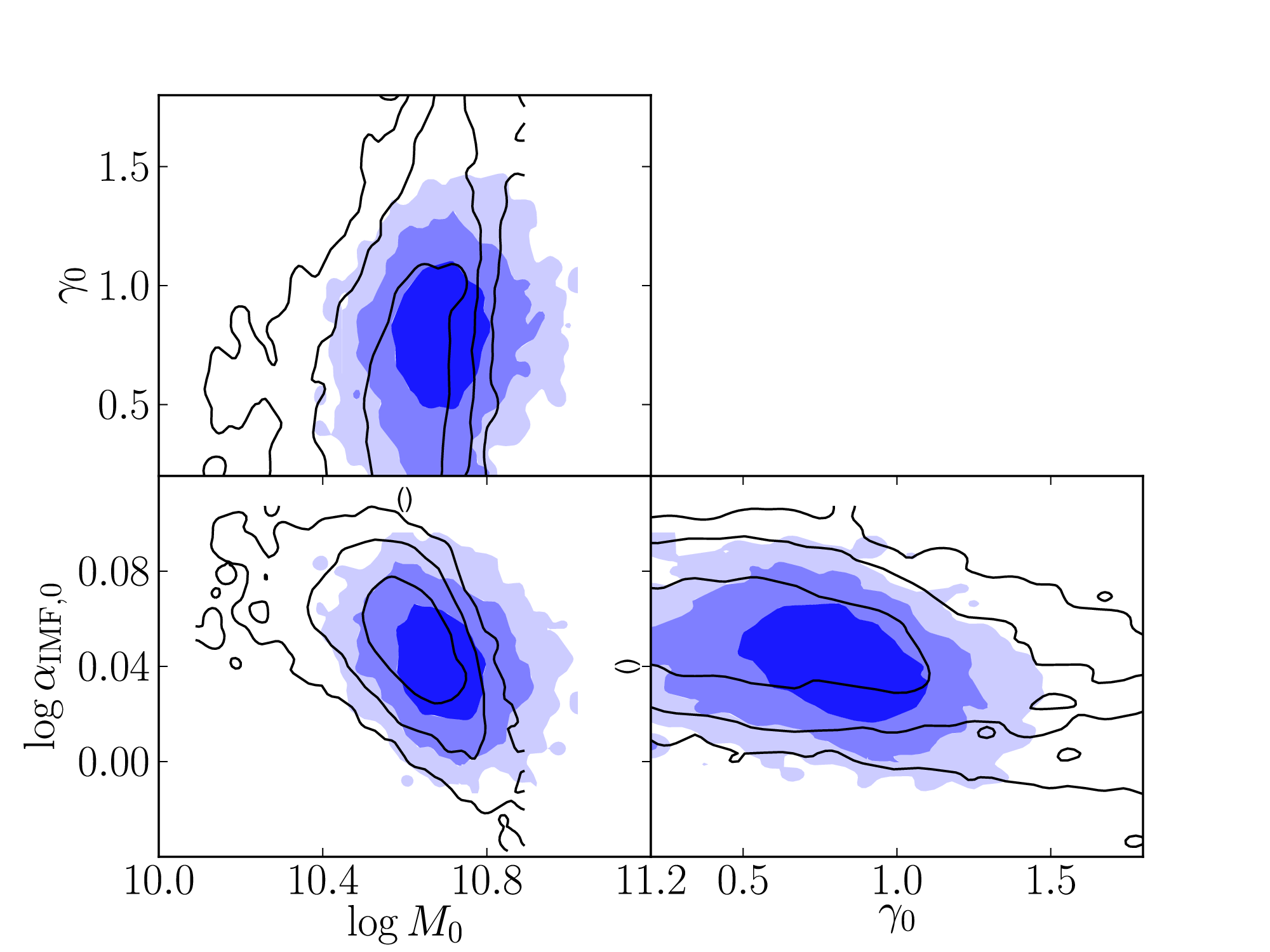}
\caption{\label{fig:degeneracy} 
Model hyper-parameters describing the average dark matter mass within 5 kpc, average dark matter slope and average IMF normalization, for galaxies at $z=0.3$, $\log{M_*} = 11.5$, $\reff = 5\rm{ kpc}$.
{\em Empty contours:} inference with no selection function term. {\em Filled contours:} including the selection function term.
}
\end{figure}
\renewcommand{\arraystretch}{1.10} 
\begin{deluxetable*}{cccl}
\tablewidth{0pt}
\tabletypesize{\small}
\tablecaption{Bayesian hierarchical inference: the hyper-parameters. gNFW model.}
\tabletypesize{\footnotesize}
\tablehead{
 & With $\mathcal{S}$ & No $\mathcal{S}$ & \colhead{Parameter description}
}
\startdata
$\log{\mpivsl2s}$ & $11.53_{-0.06}^{+0.06}$ & $11.55_{-0.06}^{+0.06}$ & Mean stellar mass at $z=0.5$, SL2S sample\\ 
$\mzsl2s$ & $0.32_{-0.31}^{+0.35}$ & $0.45_{-0.32}^{+0.33}$ & Linear dependence of mean stellar mass on redshift, SL2S sample\\ 
$\msigsl2s$ & $0.27_{-0.04}^{+0.05}$ & $0.28_{-0.04}^{+0.05}$ & Scatter in mean stellar mass, SL2S sample\\ 
$\log{\mpivslacs}$ & $11.66_{-0.03}^{+0.03}$ & $11.67_{-0.03}^{+0.03}$ & Mean stellar mass at $z=0.2$, SLACS sample\\ 
$\mzslacs$ & $2.36_{-0.43}^{+0.36}$ & $2.44_{-0.43}^{+0.35}$ & Linear dependence of mean stellar mass on redshift, SLACS sample\\ 
$\msigslacs$ & $0.23_{-0.02}^{+0.03}$ & $0.23_{-0.02}^{+0.03}$ & Scatter in mean stellar mass, SLACS sample\\ 
$\log{\rpivsl2s}$ & $0.69_{-0.04}^{+0.04}$ & $0.67_{-0.04}^{+0.04}$ & Mean effective radius at $z=0.5$, $\log{M_*} = 11.5$, SL2S sample\\ 
$\rzsl2s$ & $0.36_{-0.24}^{+0.22}$ & $0.28_{-0.20}^{+0.22}$ & Linear dependence of mean effective radius on redshift, SL2S sample\\ 
$\rmsl2s$ & $0.65_{-0.15}^{+0.16}$ & $0.69_{-0.14}^{+0.14}$ & Linear dependence of mean effective radius on stellar mass, SL2S sample\\ 
$\rsigsl2s$ & $0.18_{-0.03}^{+0.04}$ & $0.16_{-0.02}^{+0.03}$ & Scatter in mean effective radius, SL2S sample\\ 
$\log{\rpivslacs}$ & $0.70_{-0.01}^{+0.01}$ & $0.70_{-0.01}^{+0.01}$ & Mean effective radius at $z=0.2$, $\log{M_*} = 11.5$, SLACS sample\\ 
$\rzslacs$ & $0.07_{-0.17}^{+0.18}$ & $0.03_{-0.19}^{+0.18}$ & Linear dependence of mean effective radius on redshift, SLACS sample\\ 
$\rmslacs$ & $0.64_{-0.05}^{+0.05}$ & $0.63_{-0.05}^{+0.05}$ & Linear dependence of mean effective radius on stellar mass, SLACS sample\\ 
$\rsigslacs$ & $0.07_{-0.01}^{+0.01}$ & $0.07_{-0.01}^{+0.01}$ & Scatter in mean effective radius, SLACS sample\\ 
$\gammapiv$ & $0.80_{-0.22}^{+0.18}$ & $0.57_{-0.35}^{+0.41}$ & Mean $\gammadm$ at $z=0.3$, $\log{M_*} = 11.5$, $R_{\mathrm{eff}} = 5$ kpc\\ 
$\gammasig$ & $0.34_{-0.14}^{+0.27}$ & $0.69_{-0.25}^{+0.21}$ & Scatter in the $\gammadm$ distribution\\ 
$\mdmz$ & $0.86_{-0.30}^{+0.31}$ & $0.94_{-0.21}^{+0.27}$ & Linear dependence of $\mdm$ on redshift.\\ 
$\mdmm$ & $0.05_{-0.21}^{+0.22}$ & $-0.22_{-0.19}^{+0.18}$ & Linear dependence of $\mdm$ on $\log{M_*}$.\\ 
$\mdms$ & $-0.49_{-0.22}^{+0.20}$ & $-0.33_{-0.18}^{+0.17}$ & Linear dependence of $\mdm$ on $\log{\Sigma_*}$\\ 
$\log{\mdmpiv}$ & $10.69_{-0.07}^{+0.08}$ & $10.62_{-0.09}^{+0.07}$ & Mean $\mdm$ at $z=0.3$, $\log{M_*} = 11.5$, $R_{\mathrm{eff}} = 5$ kpc\\ 
$\mdmsig$ & $0.26_{-0.05}^{+0.08}$ & $0.22_{-0.04}^{+0.05}$ & Scatter in the $\mdm$ distribution\\ 
$\aimfz$ & $-0.11_{-0.09}^{+0.09}$ & $-0.06_{-0.09}^{+0.08}$ & Linear dependence of IMF normalization on redshift.\\ 
$\aimfm$ & $0.19_{-0.05}^{+0.05}$ & $0.20_{-0.04}^{+0.04}$ & Linear dependence of IMF normalization on $\log{M_*}$.\\ 
$\aimfs$ & $0.09_{-0.07}^{+0.06}$ & $0.06_{-0.07}^{+0.06}$ & Linear dependence of IMF normalization on $\log{\Sigma_*}$\\ 
$\log{\aimfpiv}$ & $0.04_{-0.02}^{+0.02}$ & $0.05_{-0.02}^{+0.02}$ & Mean IMF normalization at $z=0.3$, $\log{M_*} = 11.5$, $R_{\mathrm{eff}} = 5$ kpc\\ 
$\aimfsig$ & $0.02_{-0.01}^{+0.02}$ & $0.02_{-0.01}^{+0.02}$ & Scatter in the IMF normalization distribution\\ 
$R_{\mathrm{sel}}^{\mathrm{(SL2S)}}$ & $1.34_{-0.23}^{+0.31}$ & $\cdots$ & Mean observable Einstein radius, SL2S sample \\ 
$\sigma_{\mathrm{sel}}^{\mathrm{(SL2S)}}$ & $0.68_{-0.16}^{+0.18}$ & $\cdots$ & Dispersion in observable Einstein radius, SL2S sample \\ 
$R_{\mathrm{sel}}^{\mathrm{(SLACS)}}$ & $0.96_{-0.27}^{+0.24}$ & $\cdots$ & Mean observable Einstein radius, SLACS sample \\ 
$\sigma_{\mathrm{sel}}^{\mathrm{(SLACS)}}$ & $0.30_{-0.07}^{+0.13}$ & $\cdots$ & Dispersion in observable Einstein radius, SLACS sample \\ 

\enddata
\tablecomments{\label{tab:gnfw_inference}
Median, 16th and 84th percentile of the posterior probability distribution function of each model hyper-parameter, marginalized over the other parameters.
}
\end{deluxetable*}
The average inner dark matter slope inferred in our analysis is consistent with $\gammadm = 1$ corresponding to an NFW profile, though with a significant uncertainty: $\gammapiv = 0.80_{-0.22}^{+0.18}$.
The scatter in the slope is not well constrained and can be as large as $\sigma_\gamma \sim 0.6$.
The inference on the parameters describing the dark matter mass and IMF normalization is very similar to the NFW case: mild anticorrelation between $\mdm$ and stellar mass density and a positive correlation with redshift, no strong correlation of $\mdm$ with stellar mass, strong correlation between $\aimf$ and stellar mass. The main difference is a smaller scatter in $\mdm$ in the gNFW case.

To better illustrate the degeneracies in the model we plot in \Fref{fig:degeneracy} the projection of the posterior PDF on the parameters describing the average dark matter mass, slope and IMF normalization for galaxies at $z=0.3$, $\log{M_*}$ and $\reff=5\rm{ kpc}$.
We can see a significant degeneracy between the IMF normalization and both the dark matter mass and density slope. As discussed by \citet{Aug++10} these degeneracies are expected in a study of this nature and illustrate how independent constraints on the stellar IMF can help determine the properties of the dark matter halos of ETGs.

\begin{figure}
\includegraphics[width=\columnwidth]{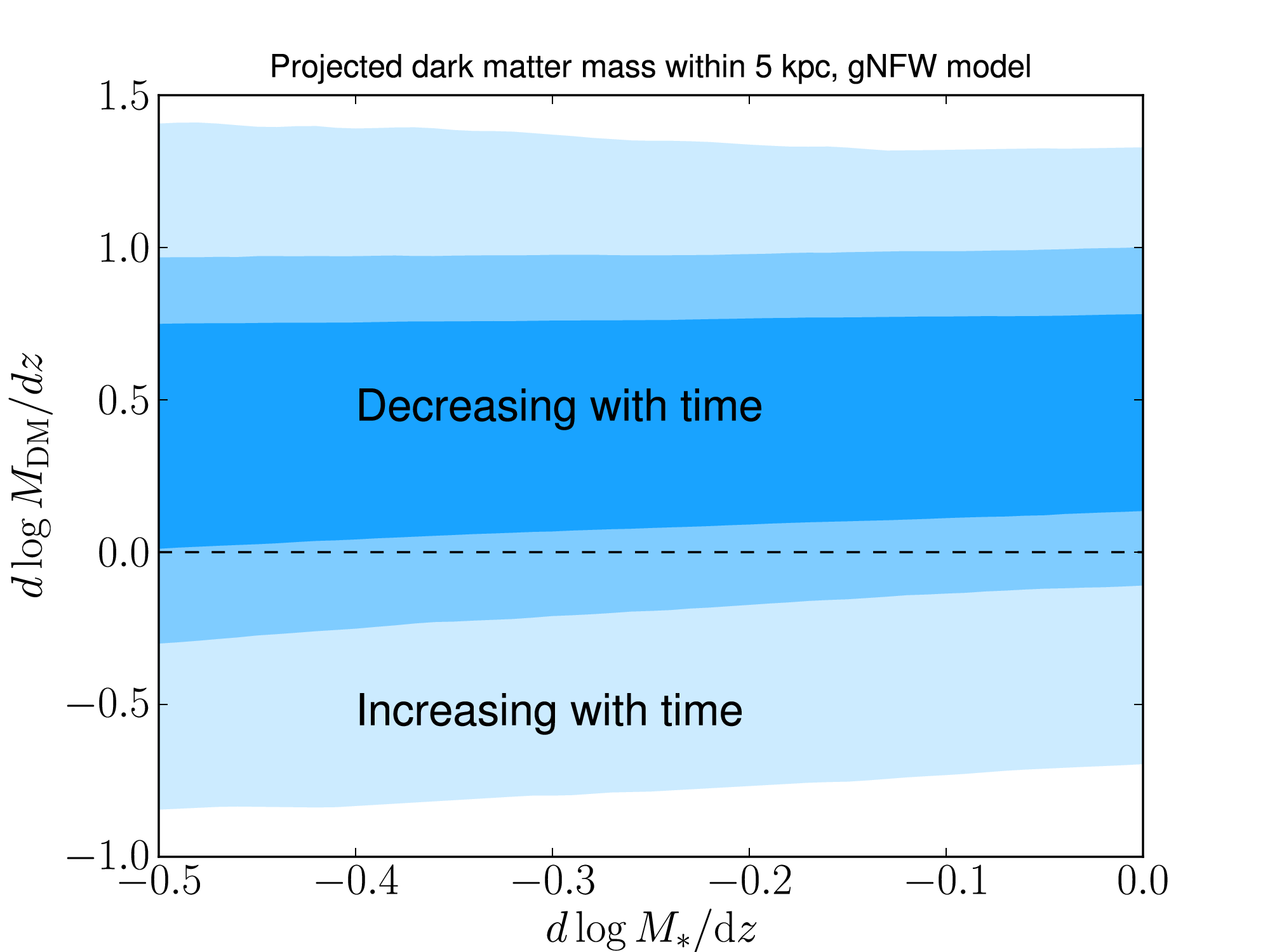}
\caption{\label{fig:mdm_lagrange_gnfw} Rate of change in projected dark matter mass within a cylinder of radius $5$ kpc along the evolutionary track of an individual galaxy, calculated from \Eref{eq:mdm_lagrange}, as a function of the growth rate in stellar mass. The dark matter halo is described with a gNFW profile. Selection effects are taken into account
}
\end{figure}

\begin{figure}
\includegraphics[width=\columnwidth]{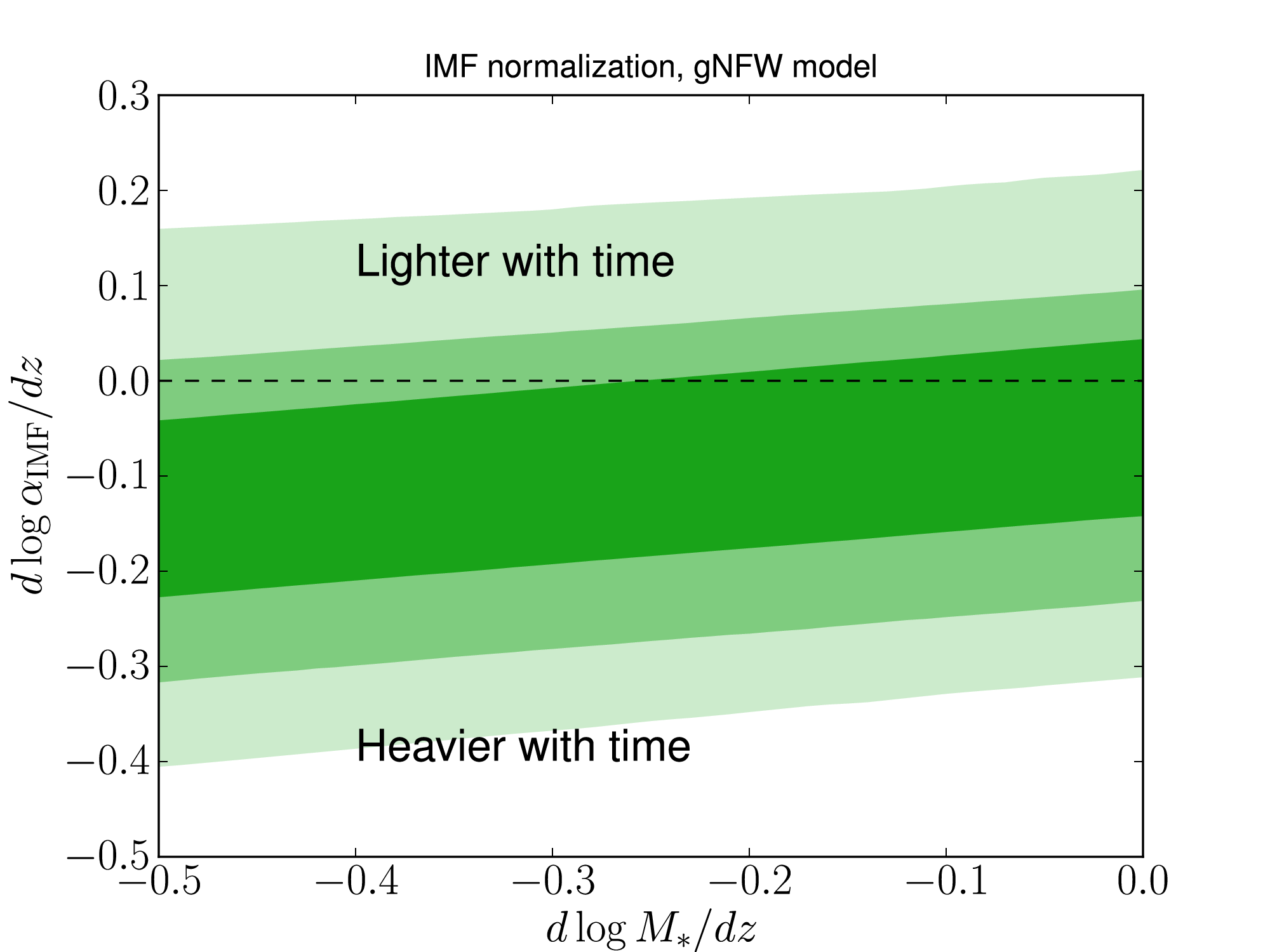}
\caption{\label{fig:imf_lagrange_gnfw} Rate of change in the IMF normalization along the evolutionary track of an individual galaxy, calculated from \Eref{eq:imf_lagrange}, as a function of the growth rate in stellar mass.
The dark matter halo is described with a gNFW profile. Selection effects are taken into account
}
\end{figure}

In continuity with the work of \Sref{ssec:nfw_lagrange}, we can calculate the rate of change of $\mdm$ and $\aimf$ along the evolutionary tracks of individual galaxies in the gNFW case.
These are plotted in \Fref{fig:mdm_lagrange_gnfw} and \Fref{fig:imf_lagrange_gnfw}.
The same operation is trivial for the dark matter slope, since we are assuming that the average slope is constant across the whole population of massive galaxies.
%
The measurements on the dark matter mass and IMF normalization are consistent with no evolution, similarly to the simpler NFW case.


\section{Discussion}\label{sect:discuss}

In Paper IV we studied the evolution of the total density profile of massive ETGs. We found that the population average slope of the density profile, $\gamma'$, increases with decreasing redshift, at fixed $M_*$ and $\reff$, and increases with $\Sigma_*$.
We also showed how $\gamma'$ stays more or less constant along the evolution of individual galaxies between $z=1$ and $z=0$.
The goal of the present paper is to understand what changes in the internal structure are responsible for the observed correlations of $\gamma'$ with $z$, $M_*$ and $\reff$.
The main steps forward in this work compared to Paper IV are 1) the use of a more physically realistic density profile, composed of a spheroid and halo instead of a single power-law component, and 2) a treatment of the lensing selection function, explicitly accounted for when deriving our results.
The latter is an important point, as it allows us to make accurate statements on the general population of massive galaxies, and not only on the population of lenses.

The analysis carried out in this paper is split into two parts: first we fix the inner slope of the dark matter halo to $\gammadm=1$, then we relax this assumption.
The inference on the population distribution of dark matter masses and stellar IMF normalization is consistent in the two cases, as the average dark matter slope inferred in \Sref{sect:gnfw} is very close to that of an NFW profile.
We found that the dark matter mass enclosed within $5$ kpc anticorrelates with the stellar mass density and positively correlates with redshift. These correlations mirror the trends of the slope of the total density profile $\gamma'$ with $\Sigma_*$ and $z$ measured in Paper IV.
At fixed redshift, galaxies with a more compact stellar distribution (larger $\Sigma_*$) tend to have smaller dark matter masses.
Stellar mass density is in turn related to the formation and evolution history. 
We know for example that minor dry mergers tend to decrease the concentration of stars by building up an extended envelope of accreted stars \citep{NJO09}.
Galaxies with a more extended stellar component then might be systems that have gone through more merger events than the average. It would then be interesting to test whether in simulations such systems are found to have larger central dark matter masses, at fixed radius, as suggested by our data.

One important point is that the inference on the evolution of the dark matter mass within $5$ kpc depends significantly on the selection function. In particular, our analysis reveals how SLACS lenses have preferentially smaller dark matter masses with respect to the population average. 
Our work is the first to explicitly fit for the selection function in deriving the properties of early-type galaxies from strong lensing measurements.
The way the selection function correction is implemented is by describing the distribution function of lenses as a product between the general distribution of massive galaxies and the probability of detecting them in lensing surveys.
The latter term is in turn the product between the lensing cross-section and an Einstein radius selection term, which describes the different probability of detecting strong lenses of different Einstein radii.
According to the works of \citet{Arn++12} and \citet{Paper0}, dedicated to the selection function of SLACS-like surveys and SL2S respectively, the Einstein radius seems to be the main quantity determining the detection probability.
Of the two terms in the selection function, the Einstein radius selection is the dominant one while the lensing cross section correction has little effect on the results of our analysis.
Strong lenses are drawn from the high mass end of the population of galaxies.
At fixed stellar mass, the difference between the distribution of strong lenses and the general distribution of galaxies is small compared to the scatter in the population.
Even though lenses with radically different density profiles can have significantly different cross-sections, as shown by \citet{MvK09}, the lensing cross-section bias is in practice small because of the small intrinsic scatter in density profile across the population of ETGs \citep[consistent with the small scatter of the mass plane,][]{Aug++10,NTB08}.

In light of this result it is important to verify the impact of the selection function on the measurement of the redshift evolution of the slope of the density profile carried out in Paper IV, which was based on the same sample of lenses used here. As we show in Appendix \ref{app:power}, the results of Paper IV are robust to selection function effects.
As a further test, we checked whether the galaxies described in our population model lie on the Fundamental Plane relation. As shown in Appendix \ref{app:fp}, that is the case.

The results presented in this work are all based on the assumption of a fixed de Vaucouleurs profile with a spatially constant mass-to-light ratio for the stellar distribution and an isotropic velocity dispersion tensor.
If any of these assumptions break down, for example with an evolving stellar profile or orbital anisotropy, then the inference might suffer from biases.
Studies of ETGs with more complex dynamical models that fit for orbital anisotropy have found no evidence for significant anisotropies \citep[e.g.][]{Cap++13}. 
It seems unlikely that allowing for anisotropy would bring significant changes to our results.
We tested for the effect of fixing the light profile to a de Vaucouleurs model by repeating the analysis of SL2S lenses with both a Hernquist \citep{Her90} and a Jaffe \citep{Jaf83} profile for the stars, and found no difference in the results.
The effect of assuming a spatially constant mass-to-light ratio can be more subtle.
In particular, if the stars accreted in merger events, which are thought to be the main drivers of the size growth of ETGs, have a lighter IMF or even a smaller mass-to-light ratio with respect to the pre-existing stellar population, then the light distribution of the stellar component will have a shallower profile than its mass distribution.
Indeed some observations suggest that the mass-to-light ratio decreases with increasing radius in early-type galaxies \citep[e.g.][]{Szo++13,Mar++14}.
In particular, \citet{Szo++13} estimate the half-mass radius to be $\sim25\%$ smaller than the half-light radius.
At fixed light profile, a galaxy with a negative gradient in the mass-to-light ratio has a steeper density profile than a model with constant $M/L$, and thus requires less stellar mass and more dark matter to produce the slope of the total density profile measured with lensing and dynamics.
If not taken into account, such a gradient in the mass-to-light ratio would then lead to an overestimate the IMF normalization and an underestimate of the dark matter mass.
More detailed data is necessary to rule out this possibility.
Nevertheless, if we repeat the analysis assuming a stellar half-mass radius $25\%$ smaller than the half-light radius for each lens, as suggested by the observations of \citet{Szo++13}, we find results consistent with the original analysis.

In this work we explored correlations between the dark matter mass and stellar IMF with redshift, stellar mass and size.
We know a more significant correlation must exist between dark matter mass and the environment of the lens, since ETGs at the center of clusters and large groups have larger projected dark matter masses than our lenses.

We leave the exploration of correlations with the environment to future work, when better data and a more extended sample of lenses will be available, covering a broader range of environments. 

\subsection{Comparison with previous works}

The inner dark matter slope of ETGs has been measured in a limited number of cases.
\citet{Son++12} measured $\gammadm=1.7\pm0.2$ for the gravitational lens SDSSJ0946$+$1006, a $z=0.222$ ETG from the SLACS sample. This value is slightly larger than the average inferred here, but is not implausible given the large scatter in $\gammadm$ of the population allowed by our data.

\citet{Gri12} found $\gammadm = 1.7\pm0.5$ for the average of the SLACS lenses assuming a Salpeter IMF, which should however be corrected to $1.40^{+0.15}_{-0.26}$ as described by \citet{D+T14}. 
In our work we let the IMF normalization be a free parameter and find a marginally shallower average dark matter slope and an IMF slightly heavier than Salpeter.
Given that most mass enclosed within the Einstein radius is stellar, a small change in the IMF can result in a significant change in the dark matter.
Indeed, if we repeat our analysis imposing a Salpeter IMF, we find much steeper dark matter slopes, consistent with the result of \citet{Gri12}.
\citet{ORF14} fitted for an average mass profile of ETG lenses in a similar way to the analysis of \citet{Gri12} but using a larger sample of lenses and including constraints from gravitational microlensing data for a few of them.
They measured the dark matter slope to be 
$\gammadm = 1.60_{-0.13}^{+0.18}$, the dark matter fraction to be around $30\%$ and find an IMF normalization slightly smaller than a Salpeter IMF. 
While dark matter fraction and IMF normalization are in good agreement with our findings, the slope of the dark matter halo measured by \citet{ORF14} is significantly larger.
Even though the lenses used in the analysis of \citet{ORF14} are for the most part the same ones used here, there are two important differences between the two works. The first difference is that \citet{ORF14} used microlensing data for a few system and no stellar kinematics information. The second difference is that we allowed for scatter in the population of galaxies, while \citet{ORF14} assumed a fixed inner slope and scaling with stellar mass of the dark matter halo, and fixed stellar IMF for all systems.
It is possible that by allowing for scatter the inference on the dark matter slope would be consistent with our results.
\citet{D+T14} find that ETGs of the mass range $\log{M_*}\sim11.5$ favor a slightly heavier than Salpeter IMF and standard NFW halos for the dark matter, in perfect agreement with our results.
\citet{Bar++13} successfully constrained the inner dark matter slope for two galaxies of the SLACS sample, thanks to a more sophisticated stellar dynamics analysis based on spatially resolved spectroscopic data.
They measured $\gammadm=0.92_{-0.64}^{+0.72}$ for SDSSJ0936+0913 and $\gammadm=0.46_{-0.30}^{+0.41}$ for SDSSJ0912+0029. 
\citet{Cap++13b} put constraints on the dark matter fractions of a large number of local ETGs from the ATLAS 3D sample finding an average fraction of $13\%$ within a sphere of radius $\reff$, corresponding to $\fdm \sim 25\%$ for an NFW profile, consistent with our results.

Concerning the IMF of ETGs and its variations with galaxy mass, a large number of works have been published in recent years.
Robust constraints on the IMF of individual systems are only available for a very limited number of objects. 
\citet{Son++12} showed that a Chabrier IMF is ruled out at 95\% confidence level in SDSSJ0946$+$1006, a much more massive ($\log{M_*}\sim11.6$) ETG.
\citet{Spi++12} found preference for a Salpeter IMF over a Chabrier IMF for a very massive lens galaxy in a group-scale halo.
\citet{Bar++13} find an IMF close to Salpeter for two SLACS lenses.
These results are consistent with our work. 
Microlensing provides an independent way to determine the absolute value of the stellar mass-to-light ratio and therefore the IMF mismatch parameter 
and the dark matter fraction. Recent works by 
\citet{ORF14} and \citet{Sch++14} find an IMF consistent with Salpeter and \citet{Jim++14} find a projected dark matter fraction consistent with our results.
A Salpeter IMF appears to be preferred over Chabrier even at $z \sim 0.8$ \citep{S+C14}, in agreement with our results.

\citet{S+L13} constrain the IMF normalization of a massive low redshift lens to be close to that of a Kroupa IMF and inconsistent with a Salpeter IMF.
While their result appears to be in tension with our model, our data allows for a certain degree of scatter in the IMF normalization and it is possible that this galaxy is just an outlier in the IMF distribution of massive ETGs, especially considering uncertainties and intrinsic scatter in the correlation between the IMF normalization and galaxy global parameters like stellar velocity dispersion.

A series of studies based on lensing and dynamics \citep{Tre++10,Aug++10b,Pos++14}, on the analysis of stellar absorption features \citep{v+C10,vDC11,vDC12,CvD12,Fer++13,LaB++13} and on spatially resolved stellar dynamics \citep{Cap++12,TRN13} have found indications for a systematic variation of the IMF with galaxy mass or velocity dispersion, with the more massive systems requiring a heavier IMF. 
Our result of an increasing IMF normalization with stellar mass further confirm the trend.
Finally, \citet{Bre++14} constrained the IMF normalization of the population of spiral galaxy bulges, with a hierarchical Bayesian inference technique similar to the one adopted in this paper, finding that the average IMF normalization must be smaller than that of a Salpeter IMF. \citet{Shu++14} find a similar result for ETGs with $\log{M_*}<10.8$. If we extrapolate our results down to the typical masses of spiral bulges, we find IMFs consistent with their results. 


\section{Summary and Conclusions}\label{sect:concl} 

We re-examined the SL2S sample of ETG lenses, extending the sample of grade A lenses and lenses usable for a joint lensing and stellar dynamics analysis with the use of key spectroscopic data recently acquired.
We then used SL2S and SLACS lenses to explore two component mass models describing the stellar spheroid and dark matter halo of massive ETGs.
We fit for the distribution function of dark matter masses, dark matter inner slopes and stellar IMF normalization across the population of massive ETGs with a Bayesian hierarchical inference method that allows for scatter in the population and takes into account the selection function, i.e. the mapping between the general population of massive galaxies and our sample of lenses.
This is the most statistically robust attempt at describing the population of ETGs with gravitational lensing data.
We found the following.
\begin{itemize}
\item The projected dark matter mass within $5$ kpc, $\mdm$, correlates with redshift and anti-correlates with stellar mass density.
The average dark matter mass for galaxies at $z=0.3$, stellar mass of $\log{M_*}=11.5$ and effective radius $\reff = 5$ kpc is $\left<\log{\mdm}\right> = 10.7\pm 0.1^{\mathrm{(stat)}} \pm 0.1^{\mathrm{(syst)}}$.
\item
SLACS lenses appear to have slightly smaller dark matter masses than the population average for galaxies of similar mass, size and redshift.
\item
The time evolution of the dark matter mass for individual objects, inferred by tracing the dark matter mass for galaxies of average mass and size at each redshift, is consistent with a mass within the inner 5 kpc that is constant with time.
Correcting for the selection function is critical for recovering this result.
\item
The average inner slope of the dark matter halos of our lenses is consistent with that of an NFW profile.
We were unable to test for correlations of the slope with redshift, stellar mass or size because the uncertainties are too large with the current data.
Spatially extended stellar kinematics data would help better constrain the dark matter slope.
\item
The IMF normalization is close to that of a Salpeter IMF and is heavier for galaxies with larger stellar mass, in agreement with previous studies.
\end{itemize}

Our finding of central dark matter content anti-correlating with stellar mass density can be interpreted as the result of more compact galaxies living in dark matter halos of smaller mass.
Stellar mass density is believed to be closely related to the assembly history of a galaxy: mergers that are predominantly dry contribute to create an extended envelope of stars, therefore galaxies with larger size might have undergone significantly more mergers with respect to more compact objects of similar mass.
Our result then seems to agree with the notion that mergers are more frequent in larger halos \citep{F+M09}, as well as with recent claims of correlation between environmental density and size of massive ETGs \citep{Coo++12b,Lan++13}.

Current and future surveys such as the Dark Energy Survey, the Large Synoptic Survey Telescope, and Euclid will provide tens of thousands of new lenses \citep{O+M10}.
Hierarchical Bayesian inference will allow to optimally combine the information from such a large number of systems and enable us to probe further the interplay between dark matter and baryons.


\acknowledgments

We thank our friends of the SLACS and SL2S collaborations for many
useful and insightful discussions over the course of the past years.
We thank V.N.~Bennert and A.~Pancoast for their help in our observational campaign.
AS acknowledges support by a UCSB Dean Graduate Fellowship.
RG acknowledges support from the Centre National des Etudes Spatiales
(CNES).
The work of PJM was supported in part by the U.S.
Department of Energy under contract number DE-AC02-76SF00515. 
TT acknowledges support from the NSF through CAREER award NSF-0642621, and from
the Packard Foundation through a Packard Research Fellowship.
CN acknowledges financial support from PRIN MIUR 2010-2011, project ``The Chemical and Dynamical Evolution of the Milky Way and Local Group Galaxies'', prot. 2010LY5N2T.
This research is based on XSHOOTER observations made with ESO Telescopes at the Paranal
Observatory under program ID 092.B-0663.
This research is based on observations obtained with MegaPrime/MegaCam, a joint project of CFHT
and CEA/DAPNIA, and with WIRCam, a joint project of CFHT, Taiwan,
Korea, Canada and France, at the Canada-France-Hawaii Telescope (CFHT) which is operated
by the National Research Council (NRC) of Canada, the Institut National des
Sciences de l'Univers of the Centre National de la Recherche Scientifique
(CNRS) of France, and the University of Hawaii. This work is based in part on
data products produced at TERAPIX and the Canadian Astronomy Data Centre.
The authors would like to thank S. Arnouts, L. Van waerbeke and G. Morrison for giving access to the WIRCam data collected in W1 and W4 as part of additional CFHT programs. We are particularly thankful to Terapix for the data reduction of this dataset.
This research is supported by NASA through Hubble Space Telescope programs
GO-10876, GO-11289, GO-11588 and in part by the National Science Foundation
under Grant No. PHY99-07949, and is based on observations made with the
NASA/ESA Hubble Space Telescope and obtained at the Space Telescope Science
Institute, which is operated by the Association of Universities for Research in
Astronomy, Inc., under NASA contract NAS 5-26555, and at the W.M. Keck
Observatory, which is operated as a scientific partnership among the California
Institute of Technology, the University of California and the National
Aeronautics and Space Administration. The Observatory was made possible by the
generous financial support of the W.M. Keck Foundation. The authors wish to
recognize and acknowledge the very significant cultural role and reverence that
the summit of Mauna Kea has always had within the indigenous Hawaiian
community.  We are most fortunate to have the opportunity to conduct
observations from this mountain.


\bibliographystyle{apj}
\bibliography{references}


\appendix
\section{Dark matter enclosed within $\reff$}\label{app:mdme}

We want to derive what our findings on the variation of $\mdm$ across the population of ETGs correspond to in terms of the projected dark matter enclosed within the effective radius, $\mdme$.
Let us derive how $\mdme$ scales with redshift, stellar mass and stellar mass density.
For simplicity we restrict ourselves to the NFW case.
For galaxies with $\reff = 5\rm{ kpc}$, $\mdm = \mdme$ by definition. Therefore for these galaxies the variation with $z$ of the dark matter mass projected within the effective radius, at fixed stellar mass and stellar mass density, is described exactly by $\mdmz$:
\begin{equation}
\frac{\partial\log{\mdme}}{\partial z} = \mdmz = 1.02_{-0.26}^{+0.32}.
\end{equation}

Let us consider the variation of $\mdme$ with stellar mass, at fixed redshift and stellar mass density. 
In order for the stellar mass density to be fixed, at a variation in stellar mass $\delta \log{M_*}$ must correspond a variation in effective radius $\delta\log{\reff} = 0.5\delta\log{M_*}$.
At fixed dark matter content, a variation in effective radius introduces a change in $\mdme$.
In particular for a galaxy with $\reff = 5\rm{ kpc}$ and an NFW dark matter halo with $r_s = 10\reff$,
\begin{equation}
\delta\log{\mdme} \approx 1.61 \delta\log{\reff}.
\end{equation}
Then, at fixed stellar mass density and redshift, the variation in $\mdme$ with stellar mass is given by the sum of a term describing the increase in halo mass, captured by the hyper-parameter $\mdmm$, and a term due to the increase in effective radius:
\begin{equation}
\frac{\partial\log{\mdme}}{\partial\log{M_*}} = \mdmm + 0.80 = 0.56 \pm 0.20.
\end{equation}

Finally a similar argument shows that, at fixed redshift and stellar mass, a variation in stellar mass density corresponds to a change in $\mdme$ given by
\begin{equation}
\frac{\partial\log{\mdme}}{\partial\log{M_*}} = \mdms - 0.80 = -1.26_{-0.33}^{+0.31}.
\end{equation}
For homologous systems, $\partial\log{\mdme}/\partial{\log{M_*}} = 1$ and $\partial\log{\mdme}/\partial{\log{\Sigma_*}} = 0$. The fact that the values we measure are inconsistent with these implies that ETGs are not homologous systems.

\section{Relation to power-law models}\label{app:power}

In Paper IV we measured the slope of the density profile and its variation across the population of strong lenses, assuming a power-law form for the density profile.
Here we are fitting a model consisting of a stellar spheroid and a dark matter halo to the same exact set of lenses.
Are the results from the two analyses consistent?
Additionally, in this work we take into account the lensing selection function. 
What would be the effect of the selection function on the analysis of Paper IV?
We can answer both these questions by generating mock samples of lenses from the population distribution inferred here, and then analyzing them with the same method of assuming power-law density profiles that we used in Paper IV.
We generated mock ensembles of 80 lenses, uniformly distributed in redshift between $z=0.1$ and $z=0.8$, with a Gaussian distribution in stellar mass centered at $\mu_* = 11.6$ and with dispersion $\sigma_* = 0.3$, values similar to the distribution of SL2S and SLACS lenses. Effective radii were drawn from a Gaussian with mean given by \Eref{eq:reff_sl2s} and dispersion $\rsigsl2s$, and dark matter masses were drawn from a Gaussian with mean given by \Eref{eq:mdm_mu} and dispersion $\mdmsig$. 
For simplicity we assumed NFW profiles for the dark matter halos, since the inference with free inner slope is consistent with that assuming NFW profiles.
The values of the hyper-parameters describing effective radius and dark matter distributions were drawn from the posterior PDF obtained from the fit described in \Sref{sect:nfw}.
For each ensemble we drew one set of hyper-parameters, and then drew the individual values of effective radii and dark matter masses.
We then simulated measurements of the density slope $\gamma'$ and added noise. This was done in Paper IV by fitting a power-law density profile to the measured central velocity dispersion and Einstein radius. 
In our case we can calculate the model velocity dispersion 
while the Einstein radius is simply set equal to the effective radius. We have shown in Paper IV that the ratio between the Einstein radius and the effective radius has little impact on the measurement of $\gamma'$.
Each mock sample is then fit with the same model for the population distribution of $\gamma'$ used in Paper IV, which consists of a Gaussian distribution with mean given by
\begin{equation}\label{eq:powerlaw}
\left<\gamma'\right> = \gamma_0' + \alpha'(z - 0.3) + \beta'(\log{M_*} - 11.5) + \xi'\log{\reff/5\rm{\,kpc}}
\end{equation}
and dispersion $\sigma_\gamma'$.
For each mock realization, we fit for the parameters of this distribution with MCMC, to give the posterior PDF for the Paper IV model parameters given the mock data. This allows us to perform the posterior predictive checks we need. For our test statistic, we predict the marginalized PDFs for the Paper IV model parameters, by considering the average of these quantities over the ensemble. Results from this exercise are reported in \Tref{table:powerlaw}.
\begin{deluxetable}{cccc}
 \tablecaption{ \label{table:powerlaw} Power-law model parameters.}
 \tablehead{
& With $\pr(\individ|\selhyperp)$ & No $\pr(\individ|\selhyperp)$ & Paper IV}
 \startdata
 $\alpha$ & $-0.30_{-0.24}^{+0.26}$ & $-0.40_{-0.18}^{+0.15}$ & $-0.31_{-0.10}^{+0.09}$ \\ 
$\beta$ & $0.35_{-0.20}^{+0.18}$ & $0.40_{-0.14}^{+0.15}$ & $0.40_{-0.15}^{+0.16}$ \\ 
$\xi$ & $-0.64_{-0.19}^{+0.19}$ & $-0.51_{-0.16}^{+0.17}$ & $-0.76_{-0.15}^{+0.15}$ \\ 
$\gamma_0$ & $1.88_{-0.08}^{+0.05}$ & $1.95_{-0.04}^{+0.04}$ & $2.08_{-0.02}^{+0.02}$ \\ 
$\sigma_{\gamma}$ & $0.15_{-0.04}^{+0.04}$ & $0.11_{-0.03}^{+0.03}$ & $0.12_{-0.02}^{+0.02}$ \\ 

 \enddata
 \tablecomments{
Fit of a Gaussian distribution in density slope with mean given by \Eref{eq:powerlaw} and dispersion $\sigma_\gamma'$ to mock populations of lenses drawn from the two component model of \Sref{sect:nfw}.
}
\end{deluxetable}
The parameters recovered in this way are well consistent with the values measured in Paper IV, with the exception of the mean density slope, $\gamma_0$.
The slope measured for mocks generated from our two component model is systematically shallower than the value measured directly on the lenses of our sample.
This discrepancy reflects the inability of reproducing relatively large values of the density slope ($\gamma' > 2.2$) with sums of de Vaucouleurs and NFW profiles, as discussed in \Sref{sect:twocomp}.
However, the key trends with $z$, $M_*$ and $\reff$ are recovered, meaning that the conclusions of Paper IV, namely that $\gamma'$ correlates with $\Sigma_*$ and anticorrelates with $z$, are perfectly consistent with the present work.
Furthermore, there is little difference between the values of the power-law parameters obtained by fitting mocks created by taking the selection function into account or not. This is an important result, as it implies that the results of Paper IV are robust with respect to selection effects.

\section{A Posterior Predictive Test}\label{app:fp}
Our hierarchical Bayesian model provides us with the the posterior probability distribution in the hyper-parameters describing the population of massive galaxies. 
One way to verify whether the inferred model is a realistic one is to draw mock observations from the posterior probability distribution and compare them with real galaxies.
In particular it is interesting to check if mock galaxies drawn from our model lie on the Fundamental Plane.
For simplicity, we consider the stellar mass Fundamental Plane \citep{HydeBernardi2009}:
\begin{equation}\label{eq:fp}
\log{\left(\frac{\reff}{\mathrm{kpc}}\right)} = a\log{\left(\frac{\sigma_0}{\mathrm{km}\,\mathrm{s}^{-1}}\right)} - 2.5b\log{\left(\frac{M_*}{2\pi\reff^2}\right)} + c,
\end{equation}
where $\sigma_0$ is the central velocity dispersion measured within an aperture $\reff/8$.
\citet{HydeBernardi2009} measured $a=1.3989$, $b=0.3164$, $c=4.4858$ from a sample of $\sim 50000$ ETGs in the SDSS. Stellar masses were obtained by \citet{Gal++05} assuming a Chabrier IMF. The observed scatter around \Eref{eq:fp} is $0.11$.

\begin{figure}
\includegraphics[width=\columnwidth]{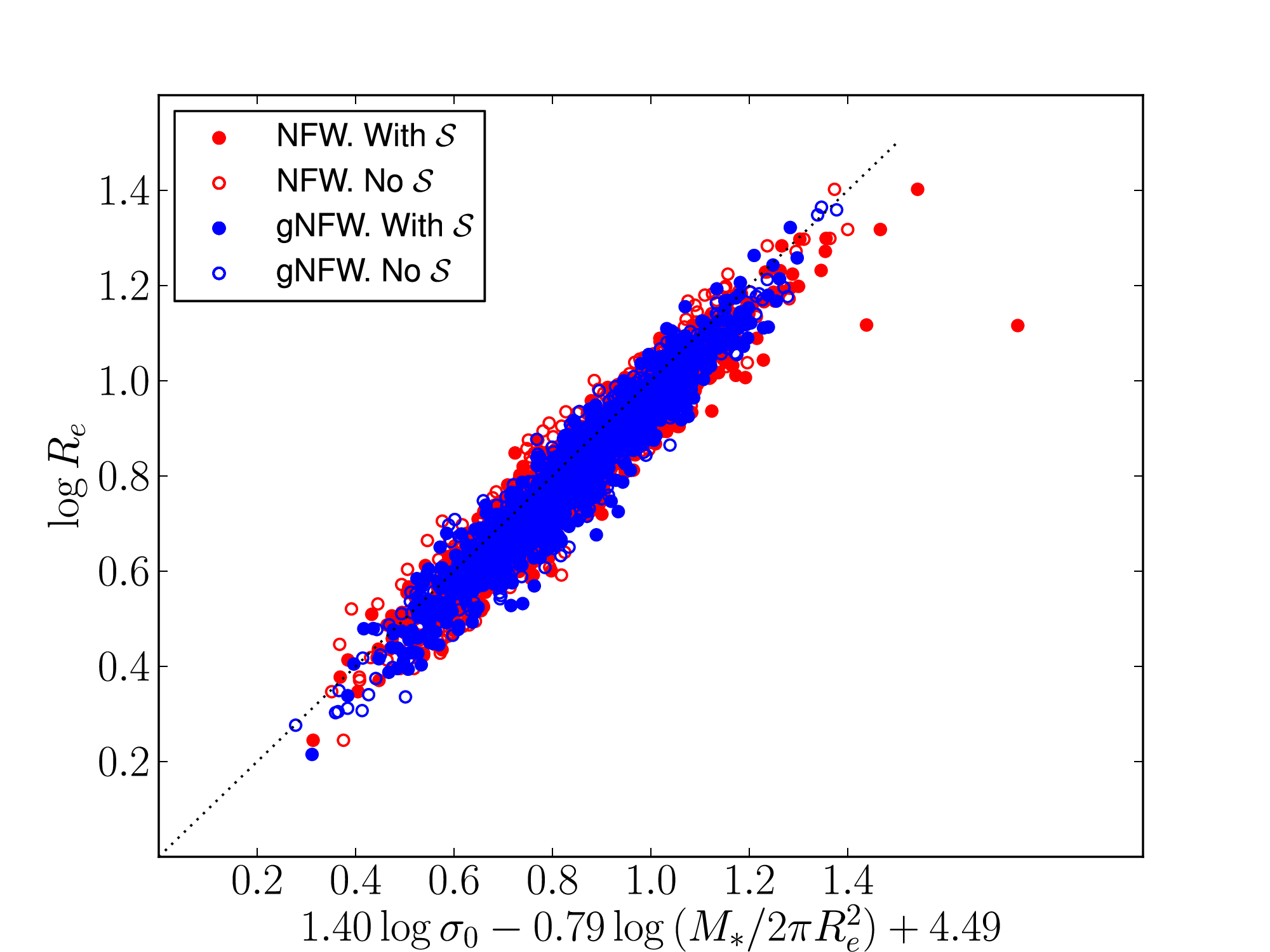}
\caption{\label{fig:fp} 
Stellar mass Fundamental Plane from mock observations generated from the posterior probability distribution function of sections 6 and 7. The coefficient of the Fundamental Plane relation are {\em not} fitted to the mock observations but are taken from the work of \citet{HydeBernardi2009}.
}
\end{figure}

In order to compare our model with the Fundamental Plane measurements we drew 1000 samples from the posterior PDF, then generated one SL2S-like galaxy for each sample and calculated the observables that enter \Eref{eq:fp}. 
Differently from the test of Appendix \ref{app:power}, we fix the galaxy redshift to $z=0.3$ for a better match with the \citet{HydeBernardi2009} sample.
Stellar population synthesis stellar masses are corrected to a Chabrier IMF for consistency.
In \Fref{fig:fp} we plot the stellar mass Fundamental Plane observed by \citet{HydeBernardi2009} together with the mock observations generated from our model.
The scatter in the mock observations is the result of both intrinsic scatter in the distribution of the parameters describing the individual galaxies and the uncertainty in the hyper-parameters.
The mock observations lie on the Fundamental Plane
Even though Fundamental Plane constraints were not explicitly used in our inference, this result shows that our model provides a correct description of the distribution in size, stellar mass and velocity dispersion.

\end{document}